\documentclass[aps,prd,preprintnumbers,superscriptaddress,showpacs,nofootinbib]{revtex4}
\usepackage[dvips]{graphicx}
\usepackage{bm,latexsym,amsmath,amssymb,amsfonts,mathrsfs}
\usepackage{color}
\usepackage{colordvi}
\input{colordvi.tex}
\usepackage{subfigure}

\newcommand{\bk}{{\bf k}}
\newcommand{\bp}{{\bf p}}

\newcommand{\rL}{{\rm L}}
\newcommand{\rM}{{\rm m}}
\begin{document}
  
\title{Halo/Galaxy Bispectrum with Primordial non-Gaussianity from integrated Perturbation Theory (iPT)}

\author{Shuichiro~Yokoyama}
\email[Email: ]{shu"at"icrr.u-tokyo.ac.jp}
\affiliation{Institute for Cosmic Ray Research, The University of Tokyo,
Kashiwa, Chiba, 277-8582, Japan}

\author{Takahiko~Matsubara}
\affiliation{Department of Physics, Nagoya University, Chikusa, Nagoya, 464-8602, Japan}
\affiliation{Kobayashi-Maskawa Institute for the Origin of Particles and the Universe, Nagoya University, Chikusa, Nagoya, 464-8602, Japan}

\author{Atsushi~Taruya}
\affiliation{Yukawa Institute for Theoretical Physics, Kyoto University, Kyoto 606-8502, Japan}
\affiliation{Research Center for the Early Universe, Graduate School of Science, The University of
Tokyo, Bunkyo-ku, Tokyo 113-0033, Japan}
\affiliation{Kavli Institute for the Physics and Mathematics of the Universe (WPI), Todai institute
for Advanced Study, University of Tokyo, Kashiwa, Chiba 277-8568, Japan}

\begin{abstract}
We derive a formula for the halo/galaxy bispectrum 
on the basis of the integrated Perturbation Theory (iPT).
In addition to the gravity-induced non-Gaussianity, 
we consider the non-Gaussianity of the primordial curvature 
perturbations, 
and investigate in detail the effect of such primordial non-Gaussianity on 
the large-scale halo/galaxy bispectrum. 
In iPT, the effects of primordial non-Gaussianity are wholly encapsulated in 
the linear (primordial) polyspectra, and we systematically calculate the 
contributions to the large-scale behaviors arising from the three types of 
primordial bispectrum (local-, equilateral-, and orthogonal-types), and 
primordial trispectrum of the local-type non-Gaussianity. 
We find that the equilateral- and orthogonal-type non-Gaussianities show 
distinct scale-dependent behaviors which can dominate the 
gravity-induced non-Gaussianity at very large scales. 
For the local-type non-Gaussianity, 
higher-order loop corrections are found to give a significantly large
contribution to the halo/galaxy bispectrum of the squeezed shape, 
and eventually dominate over the other contributions on large scales. 
A diagrammatic approach based on the iPT helps us to systematically 
investigate an impact of such higher-order contributions to the 
large-scale halo/galaxy bispectrum.
\end{abstract}

\preprint{ICRR-Report-663-2013-12}
\preprint{YITP-13-94}
\maketitle
\section{Introduction}
\label{sec:intro}

Hunting for the primordial non-Gaussianity in cosmological observations
has been a focus of attention as a big impact on the cosmology. 
In March 2013, Planck collaboration has reported tight 
constraints 
on the non-linearity parameters which characterize the amplitude of 
the deviation from pure Gaussian statistics \cite{Ade:2013ydc}. 
This result apparently implies that 
the cosmic microwave background (CMB) anisotropies measured by the Planck 
mission are very close to Gaussian, and may support
the standard scenario of structure formation that the 
large-scale structure (LSS) of the Universe 
has emerged from tiny Gaussian fluctuations. Nevertheless, the constraints 
derived from LSS observations is still limited 
(e.g., \cite{Slosar:2008hx,Giannantonio:2013uqa}), 
and at least as a cross check of
the CMB results, it is worthwhile to further investigate the 
validity of this hypothesis precisely and independently from the 
LSS observations.

It is recently known that large-scale halo/galaxy distributions 
that trace the LSS 
provide a distinct information on the primordial non-Gaussianity. 
In particular, the scale dependence of halo/galaxy bias 
has been found to be a very powerful probe to search for 
a primordial non-Gaussianity (e.g., 
\cite{Dalal:2007cu,Matarrese:2008nc,Slosar:2008hx,Giannantonio:2013uqa}). 
The most striking feature of the scale-dependent bias is that the effect 
appears even in the halo/galaxy power spectrum, and drastically change 
its shape and amplitude on large scales. This is in marked contrast to
the case of CMB observations, where the fluctuations 
are still in the linear regime, and hence
the bispectrum and other higher-order statistics 
of the CMB anisotropies are the direct probe of primordial non-Gaussianity.

In this paper, we are particularly interested in the halo/galaxy bispectrum.  
Notice that the influence of scale-dependent bias also appears in 
the halo/galaxy bispectrum and other polyspectra. Although the late-time 
gravitational evolution is known to induce the non-Gaussianity which 
inevitably dominates the primordial non-Gaussianity on small scales, 
a characteristic feature of the gravity-induced bispectrum basically 
differs from the one originating from the primordial non-Gaussianity. 
Further, due to the scale-dependent bias, the amplitude of 
halo/galaxy bispectrum may be enhanced, leading to a 
detectable signal on large scales (see, e.g., Ref.~\cite{Jeong:2009vd}). 
In Ref.~\cite{Sefusatti:2007ih,Jeong:2009vd,Sefusatti:2009qh}, 
using the peak formalism and the local bias picture, 
the authors have derived the analytic expression for halo/galaxy bispectrum. 
On the other hand, the authors of Ref.~\cite{Baldauf:2010vn} 
make use of the peak-background split picture, and 
specifically studied the halo bispectrum in the presence of primordial 
local-type non-Gaussianity,  
characterized by the two constant parameters $f_{\rm NL}$ and $g_{\rm NL}$. 
Numerical study on the halo bispectrum has been also made 
with cosmological $N$-body simulation (see, e.g., Ref.~\cite{Nishimichi:2009fs}). 
Still, however, most of works has restricted 
their attention to the local-type non-Gaussianity. 
With advent of high-precision LSS 
observations such as DES~\cite{Abbott:2005bi}, BigBOSS~\cite{Schlegel:2011zz}, 
LSST~\cite{Abell:2009aa}, EUCLID~\cite{Laureijs:2011gra}, and HSC/PFS (Sumire)~\cite{Ellis:2012rn}, it is important to give a systematic study on the 
prediction of bispectrum for various types of primordial non-Gaussianity.

In this paper, we systematically study
the effect of the primordial non-Gaussianity on
the bispectrum of the halos/galaxies, on the basis of the integrated Perturbation Theory (iPT)~\cite{Matsubara:2011ck}, 
which helps us to systematically derive a precise formula to 
connect the halo/galaxy clustering with
the initial matter density field~\cite{Matsubara:2012nc,Yokoyama:2012az}. 
With the iPT formalism, 
the non-local biasing effect can be incorporated into the statistical 
calculation of galaxies and halos in a straightforward manner. 
Further, in deriviing the effect of the primordial non-Gaussianity on 
the clustering of the halos,  
we do not need to introduce the peak-background split picture. 

In Ref. \cite{Sato:2013qfa},
the authors have compared an analytic formula for the correlation function of the halos based on the iPT formalism
with numerical $N$-body simulation and
found the agreement of those results with a high precision.
Based on this formalism, 
we compute the halo bispectrum in the presence of 
local-, equilateral-, and orthogonal-type non-Gaussianities. 
Further, in case of the local-type non-Gaussianity,  
we include the effect of the primordial trispectrum  
characterized by two non-linearity parameters, $g_{\rm NL}$ and $\tau_{\rm NL}$. 
While we mainly present the results of iPT calculation at one-loop order (i.e., next-to-leading order), the two-loop order (i.e., next-to-next-leading order) 
contributions turns out to be important in several case. We will investigate
in detail the impact of such higher-order contributions on the 
expected bispectrum signal on large scales.

This paper is organized as follows. In Sec.~\ref{sec:generalform}, we begin by presenting a general formula for the bispectrum of the biased objects, 
which includes the effect of the primordial bispectrum and trispectrum up to the one-loop order in terms of iPT. Then, in Sec.~\ref{sec:result},
we study in detail the formula for the halo bispectrum and separately consider the local-, equilateral and orthogonal-type non-Gaussianities.
In section \ref{sec:higherorder}, we investigate the contributions of the higher order loops in terms of iPT.
We discuss the comparison with the previous works where the bispectrum is obtained by other approaches
and stress  the utility of our systematic approach based on the iPT diagrammatic picture in section \ref{sec:discuss}.
We also investigate the dependences of the redshift and the mass of halos.
We devote the final section to summary.
We plot the figures of this paper with adopting the best fit cosmological parameters taken from WMAP 9-year data~\cite{Hinshaw:2012aka}.

\section{Halo/galaxy Bispectrum from integrated Perturbation Theory}
\label{sec:generalform}

In this section, we give the formula for the bispectrum of galaxies and halos, based on the integrated perturbation theory (iPT). 
As we mentioned in the introduction, 
advantage of the integrated perturbation theory (iPT) is that 
the effect of non-local biasing can be incorporated into the statistical 
calculation of galaxies and halos in a straightforward manner. 
Ref.~\cite{Sato:2013qfa} has shown that 
an analytic formula for the correlation function of the halos based on the iPT 
agrees with $N$-body simulations in a good accuracy. 
Further, we do not need to introduce the peak-background split picture 
or peak formalism
to derive the effect of the primordial non-Gaussianity in the 
clustering of the biased objects. 
Refs.~\cite{Matsubara:2012nc,Yokoyama:2012az}
have shown that 
the iPT formalism reproduces the previously known results for
the scale-dependent bias as limiting cases of a general formula,   
and thus iPT gives a more general description for the scale-dependent bias.

In Sec.~\ref{subsec:bispec_1loop},   
we first present the general expressions for the bispectrum at the one-loop 
order. We then derive the expressions of 
multi-point propagators in the large-scale limit in Sec.~\ref{subsec:Gamma_X},  
which will be the important building blocks to study the scale-dependent 
behavior of the bispectrum.

\subsection{Bispectrum at one-loop order}
\label{subsec:bispec_1loop}

We begin by defining the bispectrum of biased objects, $B_X$: 
\begin{eqnarray}
\langle \delta_X(\bk_1)\delta_X(\bk_2)\delta_X(\bk_3) \rangle
\equiv (2 \pi)^3 B_X(\bk_1,\bk_2, \bk_3) \delta^{(3)}(\bk_1+\bk_2+\bk_3)~.
\end{eqnarray}
The quantity $\delta_X$ is a Fourier transform of 
the number density field of the biased objects. 
In the iPT formalism, the multi-point propagators constitute the building blocks, and the perturbative expansion of the statistical quantities such as power spectrum and bispectrum are made with these propagators and the linear polyspectra. Denoting the $(n+1)$-point propagator of the 
biased objects by $\Gamma_X^{(n)}$, we define~\cite{Matsubara:2011ck,Bernardeau:2008fa}
\begin{eqnarray}
\Bigl\langle {\delta^n \delta_X(\bk) \over \delta \delta_\rL(\bk_1)\delta \delta_\rL(\bk_2) \cdots \delta \delta_\rL(\bk_n)}\Bigr\rangle
= (2 \pi)^{3-3n} \delta(\bk_1+\bk_2+\cdots + \bk_n) \Gamma_X^{(n)}(\bk_1,\bk_2, \cdots, \bk_n),
\end{eqnarray}
where $\delta_\rL$ represents the (initial) linear density field. 
The multi-point propagator represents the influence on $\delta_X$ due to the infinitesimal variation for the initial density fields $\delta_{\rm L}$ through the non-linear mode coupling. It 
corresponds to the summation of all the loop contributions which are
attached to each external vertex.

In order to discuss the effect of the primordial non-Gaussianity on the 
bispectrum of the biased objects, we here consider the perturbative expansion 
up to the one-loop order in iPT, which includes 
the contributions from the primordial trispectrum. 
Following the notation in the previous papers~\cite{Matsubara:2011ck,Matsubara:2012nc,Yokoyama:2012az,Matsubara:2013ofa},
the bispectrum of the biased objects is expanded as
\begin{eqnarray}
B_X(\bk_1,\bk_2,\bk_3) =
B_{\rm grav}^{\rm tree}
+B_{\rm bis}^{\rm tree}
+ B_{\rm tris}
+ B_{\rm grav}^{{\rm loop},1}
+ B_{\rm grav}^{{\rm loop},2}
+ B_{\rm bis}^{{\rm loop},1}
+ B_{\rm bis}^{{\rm loop},2}
+ B_{\rm bis}^{{\rm loop},3} +  \cdots, \nonumber
\label{eq:bispec_iPT}
\end{eqnarray} 
with
\begin{eqnarray}
B_{\rm grav}^{\rm tree} &=&
 \left[
\Gamma_X^{(1)}(\bk_1)\Gamma_X^{(1)}(\bk_2)\Gamma_X^{(2)} (-\bk_1, -\bk_2)
P_\rL (k_1) P_\rL (k_2) + 2~{\rm perms.}
\right], \cr\cr
B_{\rm bis}^{\rm tree}
& = &
 \Gamma_X^{(1)} (\bk_1) \Gamma_X^{(1)}(\bk_2) \Gamma_X^{(1)}(\bk_3)
B_\rL (\bk_1, \bk_2,\bk_3), \cr\cr
B_{\rm tris}
&=&
{1 \over 2} 
\Gamma_X^{(1)}(\bk_1)\Gamma_X^{(1)}(\bk_2) \int {d^3 p \over (2\pi)^3}
\Gamma_X^{(2)} (\bp, \bk_3 - \bp)
T_\rL (\bk_1,\bk_2,\bp,\bk_3-\bp) + 2~{\rm perms.}, \cr\cr
B_{\rm grav}^{{\rm loop},1} &=&
 \int {d^3 p \over (2 \pi)^3}
 \Gamma_X^{(2)}(\bp, \bk_1 - \bp)   \Gamma_X^{(2)}(-\bp, \bk_2 + \bp)
 \Gamma_X^{(2)}(-\bk_1+\bp, -\bk_2 - \bp)
 P_\rL(p)P_\rL(|\bk_1 - \bp|) P_\rL(|\bk_2 + \bp|) ,
\cr\cr
B_{\rm grav}^{{\rm loop},2}
&= &
 {1 \over 2}
\Gamma_X^{(1)}(\bk_1)  P_\rL (k_1)\int {d^3 p\over (2\pi)^3}
 \Gamma_X^{(2)}(\bp, \bk_2 - \bp)
\Gamma_X^{(3)}(-\bk_1, -\bp,  - \bk_2 + \bp)P_\rL(p)P_\rL(| \bk_2-\bp | ) + 5~{\rm perms.},\cr\cr
 B_{\rm bis}^{{\rm loop},1}& = & {1 \over 2} 
\Gamma_X^{(1)}(\bk_1) \Gamma_X^{(1)}(\bk_2)
\int {d^3 p\over (2\pi)^3}
\Gamma_X^{(3)}(-\bk_1,\bp,-\bk_2-\bp)P_\rL(k_1)B_\rL(\bk_2,\bp, - \bk_2-\bp)  
+ 5~{\rm perms.}, \cr\cr
B_{\rm bis}^{{\rm loop},2}
&=& {1 \over 2}
\Gamma_X^{(1)}(\bk_1)  \Gamma_X^{(2)}(-\bk_1,-\bk_2)
\int {d^3 p\over (2\pi)^3}
\Gamma_X^{(2)}(\bp, \bk_2-\bp)P_\rL(k_1)B_\rL(- \bk_2, \bp, \bk_2-\bp)  
+ 5~{\rm perms.}, \cr\cr
B_{\rm bis}^{{\rm loop},3}
&= & 
\Gamma_X^{(1)} (\bk_1) \int {d^3 p \over (2 \pi)^3}
\Gamma_X^{(2)} (\bp, \bk_2 - \bp) \Gamma_X^{(2)} (-\bp, \bk_3 + \bp) 
P_\rL (p) B_\rL (\bk_1, \bk_2+\bp, \bk_3 - \bp) + 2~{\rm perms.} .
\label{eq:ourform}
\end{eqnarray}
The functions $P_\rL$, $B_\rL$ and $T_\rL$ respectively denote 
the power-, bi- and tri-spectra of the linear density field, which are 
defined through 
\begin{eqnarray}
\langle \delta_\rL(\bk_1)\delta_\rL(\bk_2)\rangle &=& (2\pi)^3 \delta (\bk_1 + \bk_2) P_\rL(k_1), \cr\cr
\langle \delta_\rL(\bk_1)\delta_\rL(\bk_2) \delta_\rL(\bk_3)\rangle &=& (2\pi)^3 
\delta (\bk_1 + \bk_2 + \bk_3) B_\rL(k_1,k_2,k_3), \cr\cr
\langle \delta_\rL(\bk_1)\delta_\rL(\bk_2) \delta_\rL(\bk_3) \delta_\rL(\bk_4)\rangle &=& (2\pi)^3 
\delta (\bk_1 + \bk_2 + \bk_3 + \bk_4) T_\rL(k_1,k_2,k_3,k_4).
\end{eqnarray}
As we see in the above expression, in the iPT formalism,  the bispectrum of the biased objects is expressed
in terms of the primordial polyspectra as well as multi-point propagators. Hence, compared with the peak-background split picture, it is straightforward to take into account the effect of various types of the primordial non-Gaussianity whose statistical properties are encapsulated in the primordial polyspectra.
Note that the linear density field is related to the 
primordial curvature perturbations $\Phi$ through the function ${\cal M}(k)$:  
\begin{eqnarray}
\delta_\rL (k) = {\cal M}(k) \Phi(k);\quad
{\cal M}(k) = {2 \over 3} {D(z) \over D(z_*)(1+z_*)} {k^2 T(k) \over H_0^2 \Omega_{\rM 0}},
\label{eq:delta_L_Phi}
\end{eqnarray}
where $T(k)$, $D(z)$, $H_0$ and $\Omega_{\rM 0}$ are 
the transfer function, the linear growth factor, the Hubble 
parameter at present epoch, 
and the matter density parameter, respectively. 
Here $z_*$ denotes an arbitrary redshift at the matter-dominated era.
With the relation (\ref{eq:delta_L_Phi}), the linear power spectrum is given 
by
\begin{eqnarray}
P_\rL(k)=\left\{{\cal M}(k)\right\}^2 P_\Phi(k),
\end{eqnarray}
with 
\begin{eqnarray}
\langle \Phi(\bk)\Phi(\bk') \rangle = (2\pi)^3 \delta (\bk + \bk') P_\Phi(k).
\end{eqnarray}

\begin{figure}[htbp]
  \begin{center}
    \includegraphics{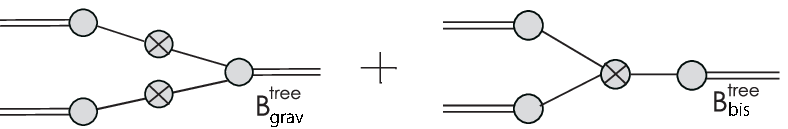}
  \end{center}
  \begin{center}
    \includegraphics{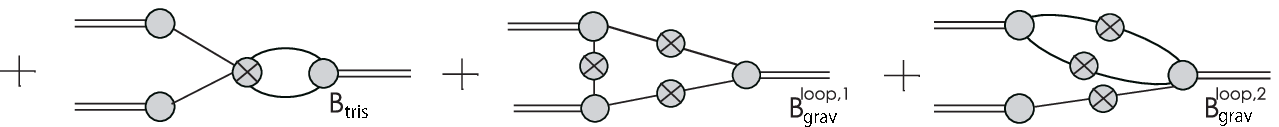}
  \end{center}
  \begin{center}
    \includegraphics{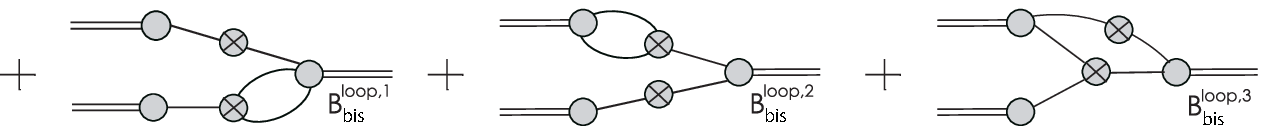}
  \end{center}
  \caption{
The diagrammatic representation for the bispectrum of 
the biased objects, $B_X$. The contributions up to the one-loop order in iPT 
are shown. A double solid line connected with a grey circle, and 
a crossed circle glued to multiple single solid lines represent 
the multi-point propagator of biased objects $\Gamma^{(n)}_X$, and 
the correlator of the initial linear density field, respectively. 
}
 \label{fig:diagram_bis.eps}
\end{figure}

In Fig. \ref{fig:diagram_bis.eps}, diagrammatic representation of each term in 
Eq.~(\ref{eq:bispec_iPT}) is shown. 
A double solid line connected with a grey circle, and 
a crossed circle glued to multiple single solid lines respectively indicate
the multi-point propagator of biased objects $\Gamma^{(n)}_X$, and 
the correlator of the initial linear density field. 
As we have mentioned, with the multi-point propagator,
all of the loop contributions which are attached to 
each external vertex are resummed, and 
the corresponding diagram does not have self-loops. 

\subsection{Multi-point propagators in the large-scale limit}
\label{subsec:Gamma_X}

The multi-point propagator $\Gamma_X^{(n)}$ 
is defined as a fully non-perturbative quantity that contains all the 
important ingredients to describe the 
non-linear gravitational evolution and galaxy/halo bias properties. 
It is therefore difficult to evaluate it rigorously,  however, 
for the large scales of our interest, perturbative treatment can work well, 
and we obtain the simplified 
expressions ~\cite{Matsubara:2011ck,Matsubara:2013ofa}. 
In particular, taking the large-scale limit
which means that the scale of interest ($\sim 1/ k_i $) is much larger than the typical scale of the formation of
the collapsed object ($\sim 1/p$ with $p$ being a variable of integration in Eq. (\ref{eq:ourform})),
we have 
\begin{eqnarray}
\Gamma_X^{(1)} (\bk) &\approx& 1 + c_1^\rL (k), \cr\cr
\Gamma_X^{(2)} (\bk_1, \bk_2) &\approx& F_2(\bk_1,\bk_2) + \left( 1+{\bk_1\cdot \bk_2 \over k_2^2}\right)c_1^\rL(\bk_1)
+ \left( 1+{\bk_1\cdot \bk_2 \over k_1^2}\right)c_1^\rL(\bk_2) + c_2^\rL(\bk_1,\bk_2),\cr\cr
\Gamma_X^{(3)} (-\bk_1, -\bp, - \bk_2 + \bp) &\approx& - {\bk_1 \cdot \bk_3 \over k_1^2} c_2^\rL(-\bp, \bp) + c_3^\rL(-\bk_1, - \bp, -\bk_2 + \bp),
\label{eq:Gamma_x_k0limit}
\end{eqnarray}
where $F_2$ is the second-order kernel of standard perturbation theory, and we have ignored the contribution of $F_3$ to $\Gamma_X^{(3)}$. The expression of the kernel $F_2$ is given by
\begin{eqnarray}
 F_2(\bk_1,\bk_2) = {10 \over 7} + \left( {k_2 \over k_1} + {k_1 \over k_2}\right) {\bk_1 \cdot \bk_2 \over k_1 k_2}
 +{4 \over 7} \left( {\bk_1 \cdot \bk_2 \over k_1 k_2}\right)^2.
\end{eqnarray}
Note that due to the symmetric property of $F_2$, we have
\begin{eqnarray}
\Gamma_X^{(2)} (-\bp, \bp) &\approx& c_2^\rL(-\bp,\bp).
\end{eqnarray}
In Eq.~(\ref{eq:Gamma_x_k0limit}), 
the quantity $c_n^\rL$ is a renormalized bias function 
defined in Lagrangian space, given by 
\begin{eqnarray}
c_n^\rL (\bk_1,\bk_2,\cdots, \bk_n) = (2\pi)^{3n} \int {d^3 k' \over (2\pi)^3} \Biggl\langle
{\delta^n \delta_X^\rL(\bk') \over \delta \delta_\rL(\bk_1)\delta \delta_\rL(\bk_2) \cdots \delta \delta_\rL(\bk_n)}
\Biggr\rangle,
\label{eq:c_n}
\end{eqnarray}
with $\delta_X^\rL$ being the number density field of biased objects defined in Lagrangian space.

So far, the derivation of the formulae above is quite general, and 
Eq. (\ref{eq:ourform}), together with the large-scale limit of 
multipoint propagators (\ref{eq:Gamma_x_k0limit}) with renormalized bias 
function (\ref{eq:c_n}),  
can be applied to any biased objects in order to study the scale-dependence 
of the bispectrum. For a quantitative investigation of the large-scale 
behaviors 
of the bispectrum, in what follows, we specifically consider the halo 
clustering, whose statistical properties are well-studied analytically  
and numerically. In the case of halos, the analytic expressions for the 
renormalized bias function $c_n^\rL$ has been already derived 
\cite{Matsubara:2011ck,Matsubara:2013ofa}, and  
we can quantitatively predict the amplitude of 
the halo bispectrum, which will be compared with $N$-body simulations. 
Since the galaxies mainly form inside the halos, for our interest of the 
large-scale limit, we expect that qualitatively 
similar features found in the case of halo bispectrum hold in the 
case of galaxy bispectrum. 

Adopting a simple model of non-local halo bias proposed by 
Ref.~\cite{Matsubara:2011ck,Matsubara:2013ofa}, 
the renormalized bias function for halos with mass $M$ is given by 
\begin{eqnarray}
c_n^\rL (\bk_1, \cdots, \bk_n) = {A_n(M) \over \delta_c^n} W(k_1;M)\cdots W(k_n;M)
+ {A_{n-1}(M) \sigma_M^n \over \delta_c^n} {d \over d \ln \sigma_M} \left[ 
{W(k_1;M) \cdots W(k_n;M) \over \sigma_M^n}
\right],
\end{eqnarray}
where $\delta_c (\simeq 1.686)$ is the so-called critical density of the spherical collapse model, 
$W(k;M)$ is the window function smoothed with the mass scale $M = 4 \pi \rho R^3 / 3$, and $\sigma_M$ is the variance of density fluctuations on the 
mass scale $M$. Here, a function $A_n(M)$ is defined by
\begin{eqnarray}
A_n(M) \equiv \sum^n_{j=0} {n! \over j!} \delta_c^j b_j^\rL(M),
\end{eqnarray}
with $b_j^\rL(M)$ being the $n$-th order scale-independent Lagrangian bias parameter which is constructed from the universal mass function as
\begin{eqnarray}
b_j^\rL(M) = \left( - \sigma_M \right)^{-n} f_{\rm MF}^{-1} {d^n \over d \nu^n} \left( f_{\rm MF}(\nu) \right).
\end{eqnarray}
Throughout the paper, we adopt the fitting formula 
by Sheth and Tormen \cite{Sheth:1999mn} for the halo mass function 
$f_{\rm MF}$, and explicitly compute the halo bispectrum:   
\begin{eqnarray}
f_{\rm MF} (\nu) = f_{\rm ST}(\nu) = A( p ) \sqrt{{2 \over \pi}}
\left[ 1 + (q\nu^2)^{-p} \right] \sqrt{q} \nu e^{-q\nu^2/2},
\end{eqnarray}
where $\nu = \delta_c / \sigma_M$, $p=0.3$, $q=0.707$ and 
the normalization factor $A( p) = \left[ 1 + \Gamma(1/2-p)/(\sqrt{\pi} 2^p) \right]^{-1}$. In Appendix \ref{sec:appendix}, 
we discuss the influence of the different choice of the mass functions on the 
final result of halo bispectrum. Note finally that 
in the large scale limit where $k_1, k_2, k_3 \to 0$, 
the window function and its derivative asymptotically approach 
$W(k_i;R) \to 1$ and
$d W(k_i;R)/d \ln \sigma_M \to 0$, and hence the renormalized bias 
function does not have significant scale-dependence.

\section{Results for each type of primordial non-Gaussianity}
\label{sec:result}

Let us now study in detail the formula for bispectrum of the biased objects 
given 
by Eq.~(\ref{eq:ourform}), especially focusing on the case of halos. 
In what follows, we separately consider the three types of primordial 
non-Gaussianity; local-, equilateral-, and orthogonal-type characterized by 
the specific shape of the bispectrum $B_\rL$ and/or trispectrum $T_\rL$.

\subsection{Local-type non-Gaussianity}
\label{sec:result_local}

In the primordial local-type non-Gaussianity, 
the bispectrum and trispectrum of linear matter density field are 
respectively characterized by 
\begin{eqnarray}
B_\rL(k_1,k_2,k_3) &=& {\cal M}(k_1){\cal M}(k_2) {\cal M}(k_3)
2 f_{\rm NL} \left[
P_\Phi(k_1)P_\Phi(k_2) + P_\Phi(k_2)P_\Phi(k_3) + P_\Phi(k_3)P_\Phi(k_1)
\right],
\end{eqnarray}
and 
\begin{eqnarray}
T_\rL(k_1,k_2,k_3,k_4) &=&  {\cal M}(k_1){\cal M}(k_2) {\cal M}(k_3){\cal M}(k_4)
\Bigl\{
6 g_{\rm NL} \left[ P_\Phi(k_1)P_\Phi(k_2)P_\Phi(k_3)  + 3~{\rm perms.} \right] \cr\cr
&&\qquad\qquad\qquad\qquad\qquad\qquad\qquad + {25 \over 9} \tau_{\rm NL} \left[ P_\Phi(k_1)P_\Phi(k_2)P_\Phi(k_{13})  + 11~{\rm perms.} \right] \Bigr\},
\end{eqnarray}
with $k_{ij} = |\bk_i + \bk_j|$. 
Here, the constant parameters, $f_{\rm NL}$, $g_{\rm NL}$ and $\tau_{\rm NL}$, are 
called non-linearity parameters. 
If one considers the case with single-sourced primordial 
curvature perturbations characterized by 
$\Phi = \Phi_G + f_{\rm NL} ( \Phi_G^2 - \langle \Phi_G^2 \rangle) + g_{\rm NL}\Phi_G^3$, 
the non-linearity parameter $\tau_{\rm NL}$ is related to  
the leading-order one $f_{\rm NL}$ 
through $\tau_{\rm NL} = 36 f_{\rm NL}^2 / 25$, which is nothing but 
the consistency relation. 
Note that this relation does not hold in general. 
In cases with multi-sourced curvature perturbations, 
we obtain the inequality, $\tau_{\rm NL} \geq 36 f_{\rm NL}^2 / 25$ 
\cite{Suyama:2007bg,Suyama:2010uj,Sugiyama:2011jt,Bramante:2011zr,Sugiyama:2012tr}. The consistency relation or inequality can be checked with 
the measurement of both the power spectrum of biased object and 
cross spectrum between the biased object and the matter density field 
\cite{Yokoyama:2012az} 
(see also Refs.\cite{Gong:2011gx,Nishimichi:2012da,Baumann:2012bc,Tseliakhovich:2010kf,Biagetti:2012xy,Smith:2010gx}).
Below, in evaluating the halo bispectrum, 
we simply assume that the consistency relation holds, 
$\tau_{\rm NL} = 36 f_{\rm NL}^2 / 25$, and present the results.

Substituting the expressions of $P_\rL$, $B_\rL$ and $T_\rL$ into 
Eq.~(\ref{eq:ourform}), each term of the bispectrum of the biased objects for the primordial local-type non-Gaussianity is evaluated as follows. Apart from 
the first term in Eq.~(\ref{eq:ourform}), the contribution $B_{\rm bis}^{\rm tree}$ becomes 
\begin{eqnarray}
B_{\rm bis}^{\rm tree} &=&
2 f_{\rm NL} \Gamma_X^{(1)}(\bk_1)\Gamma_X^{(1)}(\bk_2)\Gamma_X^{(1)}(\bk_3)
{\cal M}(k_1) {\cal M}(k_2) {\cal M} (k_3) \biggl[ P_\Phi(k_1) P_\Phi (k_2) + 2~{\rm perms.} \biggr].
\end{eqnarray}
The third term 
in Eq.~(\ref{eq:ourform}), $B_{\rm tris}$, is separately evaluated as:
\begin{eqnarray}
B_{\rm tris} &\equiv& B_{g_{\rm NL}} + B_{\tau_{\rm NL}}, \nonumber
\end{eqnarray}
with
\begin{eqnarray}
B_{g_{\rm NL}} &=& 6 g_{\rm NL}
 \Biggl\{ \Gamma_X^{(1)}(\bk_1)\Gamma_X^{(1)}(\bk_2){\cal M}(k_1){\cal M}(k_2)
 P_\Phi(k_1)P_\Phi(k_2) \cr\cr
 && \times \int {d^3 p \over (2\pi)^3} \Gamma_X^{(2)}(\bp, \bk_3 - \bp)
 {\cal M}(p){\cal M}(|\bk_3- \bp|)P_\Phi(p) \left[ 1 + {1 \over 2} \left({P_\Phi(|\bk_3 - \bp|)  \over P_\Phi(k_1)}
 + {P_\Phi(|\bk_3 - \bp|)  \over P_\Phi(k_2)} \right)
 \right] + 2~{\rm perms.} \Biggr\}, \cr\cr
 B_{\tau_{\rm NL}} &=& {25 \over 9}\tau_{\rm NL}
 \Biggl\{
 \Gamma_X^{(1)}(\bk_1)\Gamma_X^{(1)}(\bk_2){\cal M}(k_1){\cal M}(k_2)
 \left[ 
 P_\Phi(k_1) + P_\Phi(k_2)
 \right] P_\Phi(k_3) \cr\cr
 &&\qquad\qquad \qquad \times
 \int {d^3 p \over (2 \pi)^3} \Gamma_X^{(2)} (\bp, \bk_3 - \bp) {\cal M}(p){\cal M}(|\bk_3 -\bp|)
 P_\Phi(p) + 2~{\rm perms.} \Biggr\}\cr\cr
 && + {25 \over 9} \tau_{\rm NL}
 \Biggl\{ 
 \Gamma_X^{(1)}(\bk_1)\Gamma_X^{(1)}(\bk_2){\cal M}(k_1){\cal M}(k_2) P_\Phi(k_1)
P_\Phi(k_2) \cr\cr
&& \qquad\qquad\qquad \times
\int {d^3 p \over (2 \pi)^3}
  \Gamma_X^{(2)} (\bp, \bk_3 - \bp) {\cal M}(p){\cal M}(|\bk_3 -\bp|)P_\Phi(|\bk_1 + \bp|)
+ 2~{\rm perms.}\Biggr\} \cr\cr
 && + {25 \over 9} \tau_{\rm NL}
 \Biggl\{ \Biggl[
 \Gamma_X^{(1)}(\bk_1)\Gamma_X^{(1)}(\bk_2){\cal M}(k_1){\cal M}(k_2)
 P_\Phi(k_1)
 \int {d^3 p \over (2 \pi)^3} \Gamma_X^{(2)} (\bp, \bk_3 - \bp) 
 \cr\cr
 && \qquad \times
 {\cal M}(p){\cal M}(|\bk_3 -\bp|)
 P_\Phi(p) P_\Phi(|\bk_3 - \bp|)\left[ 1 + {P_\Phi(|\bk_2 + \bp|) \over  2P_\Phi(k_1)}\right] 
 + (\bk_1 \leftrightarrow \bk_2) \Biggr]
 + 2~{\rm perms.} \Biggr\}.
 \end{eqnarray}
The remaining one-loop contributions that are 
linearly proportional to $f_{\rm NL}$ are $B_{\rm bis}^{{\rm loop},1}$, 
$B_{\rm bis}^{{\rm loop},2}$, and $B_{\rm bis}^{{\rm loop},3}$, which 
can be respectively recast as
 \begin{eqnarray}
B_{\rm bis}^{{\rm loop},1} &=& 
2f_{\rm NL} \Biggl\{
 \Gamma_X^{(1)}(\bk_1)\Gamma_X^{(1)}(\bk_2)
 \Biggl[ P_\rL(k_1) {P_\rL(k_2) \over {\cal M}(k_2)} 
 \int {d^3 p \over (2 \pi)^3} \Gamma_X^{(3)}(-\bk_1, -\bp, -\bk_2 + \bp)  \cr\cr
&& \qquad\qquad\qquad \times  {\cal M}(p){\cal M}(|\bk_2 - \bp|) P_\Phi(p) 
\left( 1 + {P_\Phi(|\bk_2 - \bp|) \over 2 P_\Phi(k_2)} \right) +  
(\bk_1 \leftrightarrow \bk_2) \Biggr]
 + 2~{\rm perms.} \Biggr\},
 \cr\cr
 B_{\rm bis}^{{\rm loop},2} &=& 2 f_{\rm NL}
\Biggl\{
\Gamma_X^{(1)}(\bk_1)
\Biggl[
\Gamma_X^{(2)}(\bk_1,\bk_2)
P_\rL(k_1) {P_\rL(k_2) \over {\cal M} (k_2)} \cr\cr
&& \times
\int {d^3 p \over (2 \pi)^3}
\Gamma_X^{(2)}(\bp, \bk_2 - \bp)
{\cal M}(p){\cal M}(|\bk_2 - \bp|)P_\Phi(p)
 \left(
1 + {P_\Phi(|\bk_2 - \bp|) \over 2P_\Phi(k_2)}\right) + 
(\bk_2 \leftrightarrow \bk_3) \Biggr]
 + 2~{\rm perms.} \Biggr\}, \cr\cr
 B_{\rm bis}^{{\rm loop},3} &=& f_{\rm NL} \Biggl\{
 \Gamma_X^{(1)}(\bk_1){\cal M}(k_1)P_\Phi(k_1)
 \Biggl[
 \int {d^3 p \over (2 \pi)^3} \Gamma_X^{(2)}(\bp, \bk_2 - \bp) \Gamma_X^{(2)}( - \bp, \bk_3 + \bp)
 P_\rL(p) {\cal M}(|\bk_2 - \bp|){\cal M}(|\bk_3 + \bp|)  \cr\cr
&& \qquad\times 
\left( P_\Phi(|\bk_2 - \bp|) + P_\Phi(|\bk_3 + \bp|)  + {P_\Phi(|\bk_3 + \bp|)P_\Phi(|\bk_2 - \bp|)  \over P_\Phi(k_1)}\right) +  
(\bk_2 \leftrightarrow \bk_3) \Biggr]
 + 2~{\rm perms.} \Biggr\}.
 \end{eqnarray}

On small scales, the halo/galaxy bispectrum 
is generically dominated by the non-linearity of 
gravitational evolution, and the contribution coming 
from the higher-order loops  becomes non-negligible. 
In this respect, similar to the power spectrum case, 
large-scales are the only window where
the effect of the primordial non-Gaussianity would be significant, giving 
rise to a detectable signature on the bispectrum. 
Let us then consider the large-scale limit in which 
all of the wave numbers $k_1$, $k_2$, and $k_3$ are much smaller than the
typical scale of the biased object (halo). 
In this limit, we obtain the approximate expressions for $B_{g_{\rm NL}}$ and 
$B_{\tau_{\rm NL}}$: 
 \begin{eqnarray}
 B_{g_{\rm NL}} &\approx& 
 6 g_{\rm NL}
 \left[ \Gamma_X^{(1)}(\bk_1)\Gamma_X^{(1)}(\bk_2){\cal M}(k_1){\cal M}(k_2)
 P_\Phi(k_1)P_\Phi(k_2) \int {d^3 p \over (2\pi)^3} \Gamma_X^{(2)}(\bp,  - \bp)
P_\rL( p ) + 2~{\rm perms.} \right], \cr\cr 
 B_{\tau_{\rm NL}} &\approx&
 {25 \over 9} \tau_{\rm NL}
 \Bigl\{ \Gamma_X^{(1)}(\bk_1)\Gamma_X^{(1)}(\bk_2){\cal M}(k_1){\cal M}(k_2)\cr\cr
 && \qquad \times 
 \left[ P_\Phi(k_1) P_\Phi(k_2) + P_\Phi(k_2)P_\Phi(k_3) + P_\Phi(k_3)P_\Phi(k_1) \right]
  \int {d^3 p \over (2\pi)^3} \Gamma_X^{(2)}(\bp,  - \bp)
P_\rL( p ) + 2~{\rm perms.} \Bigr\} \cr\cr 
&& + {25 \over 9} \tau_{\rm NL}
 \Bigl\{ \Gamma_X^{(1)}(\bk_1)\Gamma_X^{(1)}(\bk_2){\cal M}(k_1){\cal M}(k_2) 
  \left[ P_\Phi(k_1) + P_\Phi(k_2) \right]
  \int {d^3 p \over (2\pi)^3} \Gamma_X^{(2)}(\bp,  - \bp)
P_\rL( p ) P_\Phi( p )+ 2~{\rm perms.} \Bigr\}.
  \end{eqnarray}
Also, the rest of the one-loop contributions is approximately described as 
 \begin{eqnarray}
 B_{\rm grav}^{{\rm loop},1} &\approx&
 \int {d^3 p \over (2\pi)^3} \left[ \Gamma_X^{(2)}(\bp, -\bp)P_\rL( p )\right]^3, \cr\cr
 B_{\rm grav}^{{\rm loop},2} &\approx& {1 \over 2}
 \left[ \Gamma_X^{(1)}(\bk_1) P_\rL(k_1)
 \int {d^3 p \over (2\pi)^3} \Gamma_X^{(2)}(\bp, -\bp)\Gamma_X^{(3)}(-\bk_1, -\bp, -\bk_2 + \bp) P_\rL( p )^2 +
 ~5~{\rm perms.}\right], \cr\cr
 B_{\rm bis}^{{\rm loop},1} &\approx& 2 f_{\rm NL}
 \left[ \Gamma_X^{(1)}(\bk_1) \Gamma_X^{(1)}(\bk_2) P_\rL(k_1) {P_\rL(k_2) \over {\cal M}(k_2)}
   \int {d^3 p \over (2\pi)^3} \Gamma_X^{(3)}(-\bk_1, -\bp, -\bk_2 + \bp) P_\rL( p ) +5~{\rm perms.} \right], \cr\cr
   B_{\rm bis}^{{\rm loop},2} &\approx& 2 f_{\rm NL}
 \left[ \Gamma_X^{(1)}(\bk_1) \Gamma_X^{(2)}(\bk_1,\bk_2) P_\rL(k_1) {P_\rL(k_2) \over {\cal M}(k_2)}
   \int {d^3 p \over (2\pi)^3} \Gamma_X^{(2)}(\bp, - \bp) P_\rL( p ) +5~{\rm perms.} \right], \cr\cr
    B_{\rm bis}^{{\rm loop},3} &\approx& 4 f_{\rm NL}
 \left\{ \Gamma_X^{(1)}(\bk_1)  {P_\rL(k_1) \over {\cal M}(k_1)}
   \int {d^3 p \over (2\pi)^3} \left[ \Gamma_X^{(2)}(\bp, - \bp) P_\rL( p )\right]^2 +2~{\rm perms.} \right\}, 
   \label{eq:lsoneloopbis}
 \end{eqnarray}
where we have used the fact that $P_\Phi ( p )  / P_\Phi (k_i) \sim k_i^3 / p^3 \to 0$.
\begin{figure}[htbp]
\begin{center}
 \includegraphics[width=150mm]{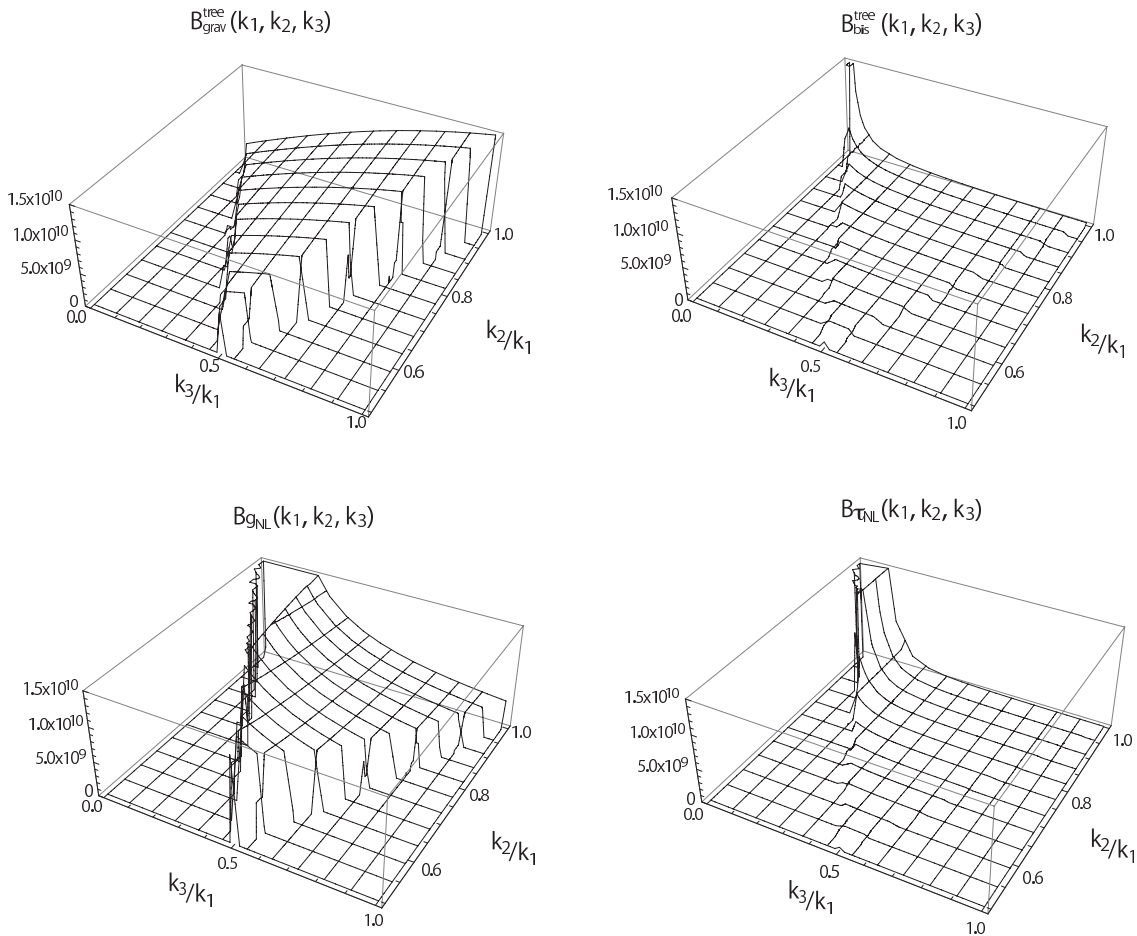}
   \end{center}
   \caption{The shapes of $B_{\rm grav}^{\rm tree}$, $B_{\rm bis}^{\rm tree}$, $B_{g_{\rm NL}}$ and $B_{\tau_{\rm NL}}$ 
   as functions of $k_3 / k_1$ and $k_2 / k_1$
   in momentum space for $k_1=0.005\,h$\,Mpc$^{-1}$.}
   \label{fig:shape.eps}
\end{figure}

As examples of the shape of each contribution in $k$-space, 
in Fig.~\ref{fig:shape.eps}, 
we plot 
$B_{\rm grav}^{\rm tree}$, $B_{\rm bis}^{\rm tree}$, $B_{g_{\rm NL}}$ and $B_{\tau_{\rm NL}}$ 
for fixed $k_1=0.005\,h$\,Mpc$^{-1}$ . 
Here, the redshift and the mass scale of halos are set to 
$z=1.0$ and $M = 5 \times 10^{13} h^{-1} M_\odot$, respectively, and we assume $k_1 \geq k_2 \geq k_3$
and $k_3 \geq k_1-k_2$ because of the triangle condition. 
In the following discussion, we fix the redshift and the mass scale of halos to be above values. 
In Discussion \ref{subsec:halo_mass_z},
we investigate the redshift and the mass dependences of 
the halo bispectrum. 
As shown in this figure, we find that the contributions of the primordial non-Gaussianity
become dominant and have large amplitudes
in the squeezed limit, while the leading-order effect of 
gravitational non-linearity appears in an equilateral shape. 
To clarify the scale-dependence of their contributions, 
let us  introduce the isosceles configuration given by 
$k \equiv k_1 = k_2 =  \alpha k_3$. A large $\alpha$ corresponds 
to the squeezed shape. Then, 
for small $k$ and large $\alpha$, 
the dominant scale-dependence of each contribution is simply given
in terms of $k$ and $\alpha$ as
\begin{eqnarray}
B_{\rm grav}^{\rm tree} \propto  k^2 \alpha^0, ~
B_{\rm bis}^{\rm tree}  \propto    k^0 \alpha^1,~
B_{g_{\rm NL}}  \propto  k^{-2} \alpha^1,~
B_{\tau_{\rm NL}} &\propto& k^{-2} \alpha^3.
\label{eq:asymptotic}
\end{eqnarray}
and
\begin{eqnarray}
B_{\rm grav}^{{\rm loop},1} \propto k^0 \alpha^0,~
B_{\rm grav}^{{\rm loop},2} \propto k^1 \alpha^0,~
B_{\rm bis}^{{\rm loop},1} \propto k^0 \alpha^1,~
B_{\rm bis}^{{\rm loop},2} \propto k^0 \alpha^1,~
B_{\rm bis}^{{\rm loop},3} \propto k^{-1} \alpha^1,
\end{eqnarray}
where we have assumed that the power spectrum of the primordial fluctuations is scale-invariant, that is, $P_\Phi (k ) \propto k^{-3} $.
From the above equations, we find that the contributions from the non-linearity of the gravitational evolution
have positive powers of $k$ and they decrease as $k$ decreases.
We also see that the contributions from the primordial bispectrum have similar $k$ and $\alpha$-dependences
except for $B_{\rm bis}^{{\rm loop},3}$. As we will see later,   $B_{\rm bis}^{{\rm loop},3}$ is suppressed on large scales
compared with other $B_{\rm bis}$ contributions, and the total contribution from the primordial bispectrum can be simply scaled as $\propto k^0 \alpha^1$. 
Such scale-dependent behaviors 
on large scales have been discussed in previous works \cite{Sefusatti:2007ih,Jeong:2009vd,Sefusatti:2009qh,Nishimichi:2009fs,Baldauf:2010vn}, 
and our result is consistent with their results. On the other hand,  
the contributions from the primordial trispectrum characterized 
by $g_{\rm NL}$ and $\tau_{\rm NL}$ (i.e., $B_{g_{\rm NL}}$ and $B_{\tau_{\rm NL}}$)  
have the $k^{-2}$-dependence, and hence we expect that 
on large scales they become dominant in the halo bispectrum.
Moreover, compared with contributions from primordial bispectrum (i.e., $B_{\rm bis}^{\rm tree}$, $B_{\rm bis}^{{\rm loop},1}$, $B_{\rm bis}^{{\rm loop},2}$, and $B_{\rm bis}^{{\rm loop},3}$) and the term $B_{g_{\rm NL}}$,
the term $B_{\tau_{\rm NL}}$ has a larger power of $\alpha$, 
and hence in the squeezed limit ($\alpha \gg 1$), this can 
dominate the halo bispectrum on large scales.

\begin{figure}[htbp]
 \begin{center}
 \includegraphics[width=150mm]{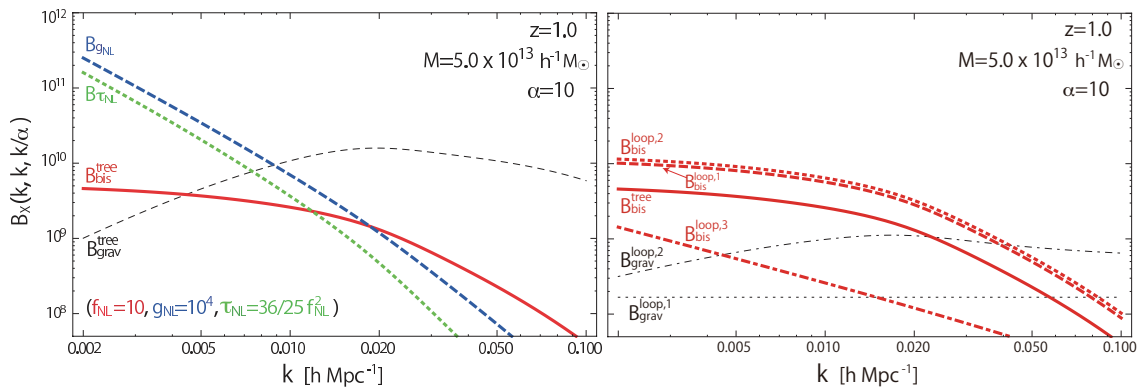}
   \end{center}
   \caption{Left panel: Contributions $B_{\rm grav}^{\rm tree}$ (black thin dashed line), 
$B_{\rm bis}^{\rm tree}$ (red thick line),
$B_{g_{\rm NL}}$ (blue thick dashed line), and
$B_{\tau_{\rm NL}}$ (green thick dotted line), as functions of the wavenumber $k$ with $\alpha=10$.
Right panel: One-loop contributions $B_{\rm grav}^{{\rm loop},1}$ (black thin dotted line), 
$B_{\rm grav}^{{\rm loop},2}$ (black thin dot-dashed line),
$B_{\rm bis}^{{\rm loop},1}$ (red thick dashed line), $B_{\rm bis}^{{\rm loop},2}$ (red thick dotted line),
$B_{\rm bis}^{{\rm loop},3}$ (red thick dot-dashed line),
and
$B_{\rm bis}^{\rm tree}$ (red thick line), as functions of the wavenumber $k$ with $\alpha=10$.
Here, we fix the redshift and the mass scale of halos to $z=1.0$ and $M = 5 \times 10^{13} h^{-1} M_\odot$, respectively.
The non-linearity parameters are given by $f_{\rm NL} = 10$, $g_{\rm NL} = 10^4$ and $\tau_{\rm NL} = 36 f_{\rm NL}^2 / 25$.}
   \label{fig:bispectrum1.eps}
\end{figure}

In left panel of Fig.~\ref{fig:bispectrum1.eps}, we plot the contributions $B_{\rm grav}^{\rm tree}$ (black thin dashed line), 
$B_{\rm bis}^{\rm tree}$ (red thick line),
$B_{g_{\rm NL}}$ (blue thick dashed line), and
$B_{\tau_{\rm NL}}$ (green thick dotted line), as functions of the wavenumber $k$ with $\alpha=10$.
The non-linearity parameters are given by $f_{\rm NL} = 10$, $g_{\rm NL} = 10^4$ and $\tau_{\rm NL} = 36 f_{\rm NL}^2 / 25$.
The asymptotic $k$-dependence expressed in Eq.~(\ref{eq:asymptotic}) is 
found to be realized at $k \lesssim 0.02\,h$\,Mpc$^{-1}$. With the 
currently allowed values of the non-linearity parameters, 
the contributions of the primordial non-Gaussianitiy can 
dominate the halo bispectrum at $k \lesssim 0.01 \,h$\,Mpc$^{-1}$. 
On the other hand, At $k \gtrsim 0.02\,h$\,Mpc$^{-1}$, the 
scale-dependence of each contribution 
gradually changes and deviates from the one in Eq.~(\ref{eq:asymptotic}). 
This is simply due to the behavior of the transfer function $T(k)$, and 
thus the approximation based on the large-scale limit would become 
invalid at $k \sim 0.1\,h$\,Mpc$^{-1}$. 

Right panel of Fig.~\ref{fig:bispectrum1.eps} shows 
the one-loop contributions, 
$B_{\rm grav}^{{\rm loop},1}$ (black thin dotted line), 
$B_{\rm grav}^{{\rm loop},2}$ (black thin dot-dashed line),
$B_{\rm bis}^{{\rm loop},1}$ (red thick dashed line), 
$B_{\rm bis}^{{\rm loop},2}$ (red thick dotted line),
$B_{\rm bis}^{{\rm loop},3}$ (red thick dot-dashed line),
and
$B_{\rm bis}^{\rm tree}$ (red thick line). The results are again plotted 
as function of the wavenumber $k$ with $\alpha=10$. 
The one-loop contributions, $B_{\rm bis}^{{\rm loop},1}$ and $B_{\rm bis}^{{\rm loop},2}$, 
dominate the tree-level contribution, $B_{\rm bis}^{\rm tree}$. 
This fact has been also addressed 
by Jeong and Komatsu (2009) \cite{Jeong:2009vd}, 
who computed the halo bispectrum on the basis of the peak formalism, 
taking account of the contribution from the
primordial trispectrum.

The reason why the contributions $B_{\rm bis}^{{\rm loop},1}$ and 
$B_{\rm bis}^{{\rm loop},2}$ become larger than the other one-loop 
contributions may be explained from
the diagrams shown in Fig.~\ref{fig:diagram_bis.eps}. Although 
all the diagrams in iPT are irreducible, 
the diagram of $B_{\rm bis}^{{\rm loop},3}$ graphically looks 
similar to $B_{\rm bis}^{\rm tree}$, and connecting two of the three grey circles with linear power spectrum gives $B_{\rm bis}^{{\rm loop},3}$. In similar way, 
$B_{\rm grav}^{{\rm loop},1}$ and $B_{\rm grav}^{{\rm loop},2}$ can be constructed 
from the tree diagram $B_{\rm grav}^{\rm tree}$ by adding a power spectrum. We 
may call them {\it decomposable} diagrams. 
On the other 
hand, the diagrams $B_{\rm bis}^{{\rm loop},1}$ and $B_{\rm bis}^{{\rm loop},2}$ 
can not 
be constructed from the tree diagrams by simply adding a power spectrum. 
We may call such kind of contributions {\it un-decomposable} diagrams. 
The un-decomposable diagrams in nature involve higher-order correlators of the initial linear density field, and these correlators form a specific type of loops by glueing some of the legs (indicated by solid lines) to a multi-point propagator. In the large-scale limit, their loop integral can be dominated by the contributions of the squeezed limit of the higher-order correlators. Since the local-type non-Gaussianity is known to produce a large primordial bispectrum in the squeezed limit, the un-decomposable diagram can potentially give a significantly large contribution to the halo bispectrum. Indeed, the expressions of  
$B_{\rm bis}^{{\rm loop},1}$ and $B_{\rm bis}^{{\rm loop},2}$ in the large-scale limit, given in Eq.~(\ref{eq:lsoneloopbis}), are mostly dominated by the squeezed limit of $B_{\rm L}$, and the effect of a strong coupling between short- and long-modes results in both the scale-dependence proportional to $\mathcal{M}(k)^{-1} \propto k^{-2}$ and the integral retaining the short-mode contribution that produces a very large amplitude.

\begin{figure}[htbp]
 \begin{center}
  \includegraphics[width=80mm]{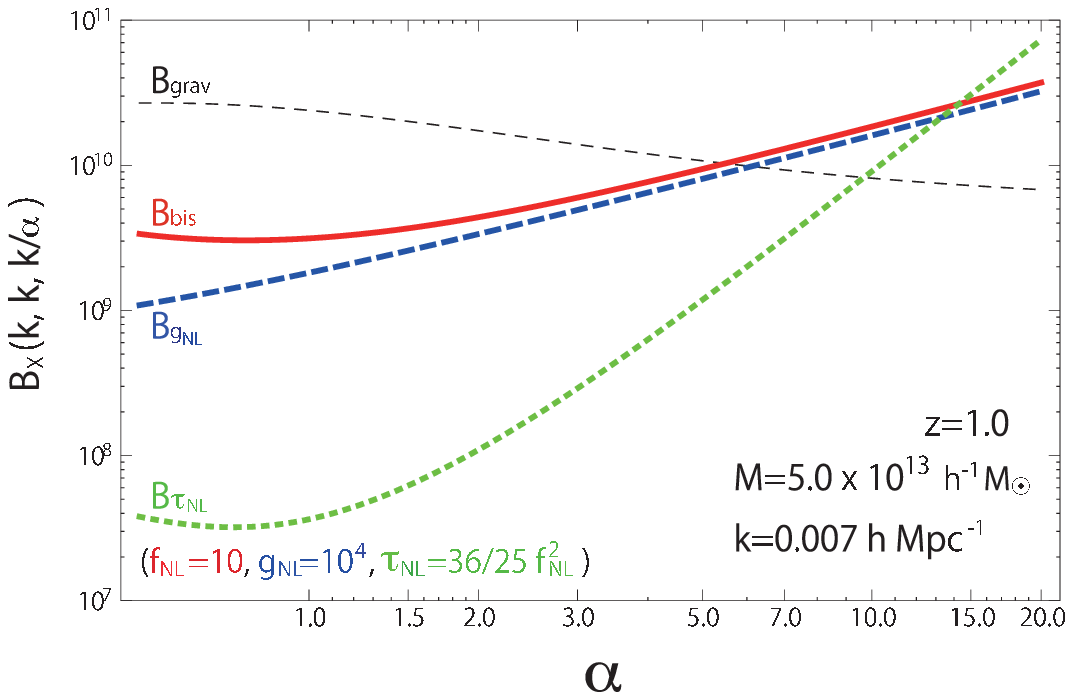}
   \end{center}
   \caption{Contributions $B_{\rm grav} (=B_{\rm grav}^{\rm tree} + B_{\rm grav}^{{\rm loop},1}+B_{\rm grav}^{{\rm loop},2})$ (black thin dashed line), 
$B_{\rm bis} (= B_{\rm bis}^{\rm tree} + B_{\rm bis}^{{\rm loop},1}+B_{\rm bis}^{{\rm loop},2}+ B_{\rm bis}^{{\rm loop},3})$ (red thick line),
$B_{g_{\rm NL}}$ (blue thick line), and
$B_{\tau_{\rm NL}}$ (green thick line), as functions of $\alpha$ with $k= 0.007\, h\, {\rm Mpc}^{-1}$.
Here, we fix that the redshift $z=1.0$ and the mass scale of halos $M = 5 \times 10^{13} h^{-1} M_\odot$.
The non-linearity parameters are given by $f_{\rm NL} = 10$, $g_{\rm NL} = 10^4$ and $\tau_{\rm NL} = 36 f_{\rm NL}^2 / 25$.}
   \label{fig:bispectrum2.eps}
\end{figure}

Finally, in Fig.~\ref{fig:bispectrum2.eps}, we compare the total 
contribution from 
the gravitational non-linearity with those from primordial non-Gaussianity. 
Plotted results are 
the contributions $B_{\rm grav} (=B_{\rm grav}^{\rm tree} + B_{\rm grav}^{{\rm loop},1}+B_{\rm grav}^{{\rm loop},2})$ (black thin dashed line), 
$B_{\rm bis} (= B_{\rm bis}^{\rm tree} + B_{\rm bis}^{{\rm loop},1}+B_{\rm bis}^{{\rm loop},2}+ B_{\rm bis}^{{\rm loop},3})$ (red thick line),
$B_{g_{\rm NL}}$ (blue thick line), and
$B_{\tau_{\rm NL}}$ (green thick line), as functions of $\alpha$, fixing the
wavenumber to $k=0.007\,h\,{\rm Mpc}^{-1}$. 
As we saw before, the contributions from the primordial non-Gaussianity become 
large as increasing $\alpha$, and eventually dominate the halo bispectrum.  
In particular, in the squeezed limit ($\alpha \gg 1$),  
the term $B_{\tau_{\rm NL}}$ is found to exceed other non-Gaussian contributions. 
%

\subsection{Equilateral-type non-Gaussianity}

Let us next consider the 
equilateral-type non-Gaussianity, in which the bispectrum of linear density field $B_\rL$ is 
given by  
\begin{eqnarray}
B_\rL ({\bf k}_1, {\bf k}_2, {\bf k}_3) &=& 6 f_{\rm NL}^{\rm equil}{\cal M}(k_1){\cal M}(k_2){\cal M}(k_3)
\left[
- \left(P_\Phi(k_1)P_\Phi(k_2) + 2~{\rm perms.}\right) \right. \cr\cr
&&\left.
- 2 P_\Phi(k_1)^{2/3}P_\Phi(k_2)^{2/3} P_\Phi(k_3)^{2/3}
+ \left(P_\Phi(k_1)^{1/3}P_\Phi(k_2)^{2/3}P_\Phi(k_3) + 5~{\rm perms.}\right)
\right].
\label{eq:eqbi}
\end{eqnarray}
Here, we do not consider the contribution from the "equilateral"-trispectrum, 
because its scale dependence strongly depends on the models of
generating primordial non-Gaussianity. Also, its exact form 
is much complicated compared to that of the local-type non-Gaussianity.
In fact, there are several works on the estimator for the 
trispectrum for the CMB temperature fluctuations in the models producing 
the equilateral-type bispectrum~\cite{Mizuno:2010by,Izumi:2011di}. 
We leave the discussion on the contribution from the equilateral-trispectrum 
to future work.

The tree-level contribution from the primordial bispectrum, $B_{\rm bis}^{\rm tree}$, becomes 
\begin{eqnarray}
B_{\rm bis}^{\rm tree} &=& 6 f_{\rm NL}^{\rm equil}
\Gamma_X^{(1)}(\bk_1)\Gamma_X^{(1)}(\bk_2)\Gamma_X^{(1)}(\bk_3)
{\cal M}(k_1){\cal M}(k_2){\cal M}(k_3)
\left[
- \left(P_\Phi(k_1)P_\Phi(k_2) + 2~{\rm perms.}\right) \right. \cr\cr
&&\left.
- 2 P_\Phi(k_1)^{2/3}P_\Phi(k_2)^{2/3} P_\Phi(k_3)^{2/3}
+ \left(P_\Phi(k_1)^{1/3}P_\Phi(k_2)^{2/3}P_\Phi(k_3) + 5~{\rm perms.}\right)
\right].
\end{eqnarray}
On the other hand, the one-loop contributions, $B_{\rm bis}^{{\rm loop},1}$ and $B_{\rm bis}^{{\rm loop},2}$ include the primordial bispectrum of the specific 
configuration, 
$B_\rL(\bk_i, -\bp, -\bk_i + \bp)$. In the large-scale limit $k_i/p \ll 1$, the leading-order behavior becomes
\begin{eqnarray}
B_\rL(\bk_i, - \bp, -\bk_i + \bp) \approx 12 f_{\rm NL}^{\rm equil}  {P_\rL(k_i) \over {\cal M}(k_i)}P_\rL( p ) \left( {k_i \over p} \right)^2
 \left[ \left( {k_i \over p} \right)^{- {2 \over 3} (n_s-1) } - \left( {n_s -4 \over 3} \right)^2 \mu_i^2 \right].
\end{eqnarray}
Here  
$\mu_i \equiv  { \bk_i \cdot \bp \over k_i p}$ and
we assume the power-law spectrum, 
$P_\Phi(k) \propto k^{n_s - 4}$. 
Also, $B_{\rm bis}^{{\rm loop},3}$ has the primordial bispectrum of the configuration, $B_\rL(\bk_1, \bk_2 + \bp, \bk_3 -  \bp)$, which can be reduced to
\begin{eqnarray}
B_\rL(\bk_1, \bk_2 + \bp, \bk_3 -  \bp) \approx 12 f_{\rm NL}^{\rm equil}
 {P_\rL(k_1) \over {\cal M}(k_1)}P_\rL( p ) \left( {k_1 \over p} \right)^2
  \left[ \left( {k_1 \over p} \right)^{- {2 \over 3} (n_s-1) } - \left( {n_s -4 \over 3} \right)^2 \left({k_2^2 \mu_2^2 + k_3^2 \mu_3^2 \over k_1^2}\right) \right].
\end{eqnarray}
Then, we obtain the scale-dependence for the halo bispectrum induced from the equilateral primordial non-Gaussianity.  
In the large scale limit, we have 
\begin{eqnarray}
B_{\rm bis}^{\rm tree} \propto k^0 \left( {2 \alpha - 1 \over \alpha^2}\right),~
B_{\rm bis}^{{\rm loop},1} \propto k^2 \alpha^0,~
B_{\rm bis}^{{\rm loop},2} \propto k^2 \alpha^0,~
B_{\rm bis}^{{\rm loop},3} \propto k^{1} \alpha^1.
\end{eqnarray}
Here we simply assume the scale-invariant 
primordial power spectrum, that is, $n_s = 1$.
Because of the positive powers of $k$,
the one-loop contributions from the equilateral-type bispectrum are all 
suppressed on the large scales, 
compared to the tree-level contribution.
\begin{figure}[htbp]
 \begin{center}
  \includegraphics[width=150mm]{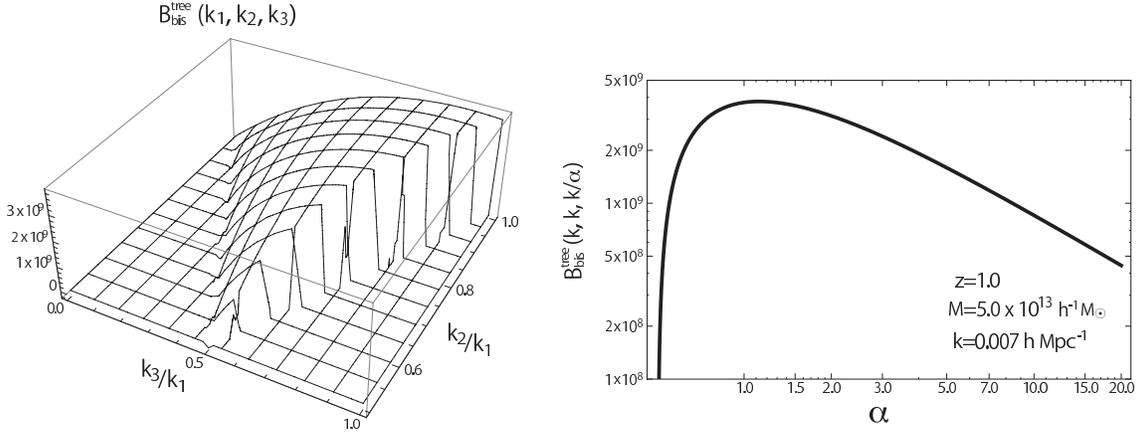}
   \end{center}
   \caption{$B_{\rm bis}^{\rm tree}(k_1,k_2,k_3)$ 
   as function of $k_3 /k_1$ and $k_2 / k_1$ for $k_1 = 0.007\, h\, {\rm Mpc}^{-1}$ (left panel), and
  $B_{\rm bis}^{\rm tree}(k_1,k_2,k_3)$ as a function of $\alpha$  with fixing $k = 0.007\, h\, {\rm Mpc}^{-1}$
   (right panel)
   for equilateral-type primordial non-Gaussianity. We take $f_{\rm NL}^{\rm equil} = 100$.}
   \label{fig:eq_alpha.eps}
\end{figure}

Left panel of Fig.~\ref{fig:eq_alpha.eps} shows the shape of $B_{\rm bis}^{\rm tree}(k_1,k_2,k_3)$
in the momentum space, fixing wavenumber $k_1$ to 
$0.007\, h\,{\rm Mpc}^{-1}$, and we assume $k_1 \geq k_2 \geq k_3$
and $k_3 \geq k_1-k_2$ because of the triangle condition.
Right panel of Fig.~\ref{fig:eq_alpha.eps} shows the $\alpha$-dependence of $B_{\rm bis}^{\rm tree}(k_1,k_2,k_3)$, for
isosceles configuration as $k_1 = k_2 = \alpha k_3 = k$, and we 
fix the wavenumber $k$ to $0.007\, h\, {\rm Mpc}^{-1}$ and 
take $f_{\rm NL}^{\rm equil} = 100$, 
 which is almost the $2$-$\sigma$ upper bound 
obtained by Planck collaboration \cite{Ade:2013ydc}.
The amplitude of  $B_{\rm bis}^{\rm tree}$ 
has a peak around $\alpha = 1.0$, indicating 
that the halo bispectrum has equilateral shape in $k_1,k_2,k_3$ space. 
Since the contribution from the non-linearity of 
the gravitational evolution $B_{\rm grav}^{\rm tree}$ also has 
a peak around $\alpha = 1.0$ (see Fig.~\ref{fig:shape.eps}), 
it seems difficult to 
distinguish between $B_{\rm bis}^{\rm tree}$ and $B_{\rm grav}^{\rm tree}$
in the case of equilateral-type non-Gaussianity.

\begin{figure}[htbp]
 \begin{center}
  \includegraphics[width=80mm]{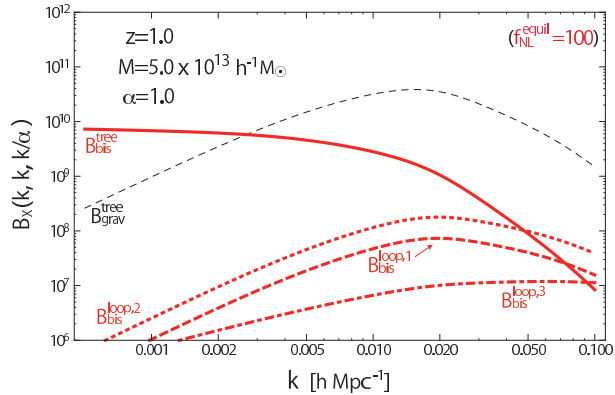}
   \end{center}
   \caption{$B_{\rm grav}^{\rm tree}$ (black dashed line), $B_{\rm bis}^{\rm tree}$ (red thick line), 
   $B_{\rm bis}^{{\rm loop},1}$ (red dashed thick line), $B_{\rm bis}^{{\rm loop},2}$ (red dotted thick line),
   and  $B_{\rm bis}^{{\rm loop},3}$ (red dotted-dashed thick line)
   as a function of $k$  with fixing $\alpha=1.0$
   for equilateral-type primordial non-Gaussianity. We take $f_{\rm NL}^{\rm equil} = 100$. }
   \label{fig:eq_k.eps}
\end{figure}

However, the $k$-dependence of $B_{\rm bis}^{\rm tree}$ and $B_{\rm grav}^{\rm tree}$ 
has very distinct feature, as shown in Fig. \ref{fig:eq_k.eps}. 
Here, we plot $B_{\rm grav}^{\rm tree}$ (black dashed line) 
and $B_{\rm bis}^{\rm tree}$ (red thick line) as a function of $k$, fixing $\alpha$ to unity and taking $f_{\rm NL}^{\rm equil} = 100$.
For the non-linearity parameter consistent with the current observational limit,
we find that $B_{\rm bis}^{\rm tree}$ dominates $B_{\rm grav}^{\rm tree}$ at $k \lesssim 0.003\, h\, {\rm Mpc}^{-1}$. 
Note that the detectability of the equilateral primordial no-Gaussianity from 
galaxy survey has been discussed in Ref.~\cite{Sefusatti:2007ih}, which also 
found the similar scale-dependent behavior. 
The authors of Ref.~\cite{Sefusatti:2007ih} computed 
the galaxy bispectrum based on the local bias ansatz. 
At the tree level, 
the difference between our formula and
the expression in Ref.~\cite{Sefusatti:2007ih} 
appears in the scale dependence of the term, $(1 + {\bk_1 \cdot \bk_2}/k_1^2)c_2^\rL$ 
included in $\Gamma_X^{(2)}(\bk_1,\bk_2)$.
This basically comes from the bias prescription which we adopted here \cite{Matsubara:2011ck,Matsubara:2013ofa}, and reflects the non-local nature of the halo bias, but it does not produce much difference in the halo bispectrum on large scales.

In Fig.~\ref{fig:eq_k.eps}, we also plot 
the one-loop contributions for the equilateral-type; 
$B_{\rm grav}^{\rm tree}$ (black dashed line), $B_{\rm bis}^{\rm tree}$ (red thick line), 
   $B_{\rm bis}^{{\rm loop},1}$ (red dashed thick line), $B_{\rm bis}^{{\rm loop},2}$ (red dotted thick line), and  $B_{\rm bis}^{{\rm loop},3}$ (red dotted-dashed thick line). 
The one-loop contributions are all suppressed on large scales as mentioned before. 
In Ref.~\cite{Sefusatti:2009qh}, the one-loop corrections 
have been also discussed. 
The author computed the one-loop contributions with the 
Eulerian local bias prescription, 
and mentioned that one of the loop contributions is not negligible even 
for the equilateral-type non-Gaussianity. 
In the diagrammatic picture, this corresponds to the loop 
attached to an external vertex. 
In our formalism on the basis of iPT, however,  such loops are already included 
in the multi-point propagator as a result of the resummation.   
Hence such higher-order loop corrections do not appear in our formula. 

\subsection{Orthogonal-type primordial non-Gaussianity}
\label{sec:orthogonal}

As the third type of primordial non-Gaussianity, we consider the 
orthogonal-type non-Gaussianity. The bispectrum of the 
orthogonal type is defined as
\begin{eqnarray}
B_\rL ({\bf k}_1, {\bf k}_2, {\bf k}_3) &=& 6 f_{\rm NL}^{\rm orth}{\cal M}(k_1){\cal M}(k_2){\cal M}(k_3)
\left[
-3 \left(P_\Phi(k_1)P_\Phi(k_2) + 2~{\rm perms.}\right) \right. \cr\cr
&&\left.
- 8 P_\Phi(k_1)^{2/3}P_\Phi(k_2)^{2/3} P_\Phi(k_3)^{2/3}
+3 \left(P_\Phi(k_1)^{1/3}P_\Phi(k_2)^{2/3}P_\Phi(k_3) + 5~{\rm perms.}\right)
\right]. 
\label{eq:eqbi}
\end{eqnarray}
Then we obtain 
\begin{eqnarray}
B_{\rm bis}^{\rm tree} &=& 6 f_{\rm NL}^{\rm orth}
\Gamma_X^{(1)}(\bk_1)\Gamma_X^{(1)}(\bk_2)\Gamma_X^{(1)}(\bk_3)
{\cal M}(k_1){\cal M}(k_2){\cal M}(k_3)
\left[
- 3\left(P_\Phi(k_1)P_\Phi(k_2) + 2~{\rm perms.}\right) \right. \cr\cr
&&\left.
- 8 P_\Phi(k_1)^{2/3}P_\Phi(k_2)^{2/3} P_\Phi(k_3)^{2/3}
+3 \left(P_\Phi(k_1)^{1/3}P_\Phi(k_2)^{2/3}P_\Phi(k_3) + 5~{\rm perms.}\right)
\right].
\end{eqnarray}
\begin{figure}[htbp]
 \begin{center}
  \includegraphics[width=150mm]{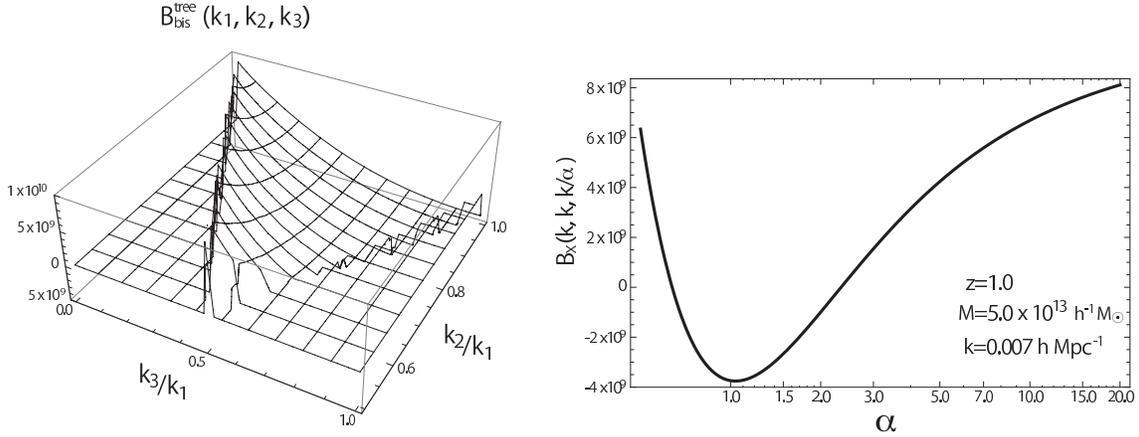}
   \end{center}
   \caption{Left panel; $B_{\rm bis}^{\rm tree}(k_1,k_2,k_3)$ 
   as function of $k_3 /k_1$ and $k_2 / k_1$ for $k_1 = 0.007 h {\rm Mpc}^{-1}$,
   Right panel; $B_{\rm bis}^{\rm tree}(k_1,k_2,k_3)$ as a function of $\alpha$  with fixing $k = 0.007 h {\rm Mpc}^{-1}$, 
   for orthogonal-type primordial non-Gaussianity. We take $f_{\rm NL}^{\rm orth} = -100$.}
   \label{fig:ort_alpha.eps}
\end{figure}
In Fig.~\ref{fig:ort_alpha.eps}, shape of the bispectrum 
$B_{\rm bis}^{\rm tree}(k_1,k_2,k_3)$ (left) and 
its $\alpha$-dependence (right) are shown.  
For the parameter $f_{\rm NL}^{\rm orth}$, we adopt  $f_{\rm NL}^{\rm orth}= -100$, 
consistent with the Planck results at $2$-$\sigma$ level \cite{Ade:2013ydc}. 
Right panel shows as increasing $\alpha$, 
$B_{\rm bis}^{\rm tree}$ starts to decrease and has a negative 
peak around $\alpha = 1.0$, and then turns to  
increase with positive value toward the squeezed limit 
($\alpha\gg1$). This is in marked contrast to 
the behaviors in other types of the non-Gaussianity, 
and may be a very important clue 
to separately detect the orthogonal-type non-Gaussianity from the 
halo bispectrum. 
\begin{figure}[htbp]
 \begin{center}
  \includegraphics[width=80mm]{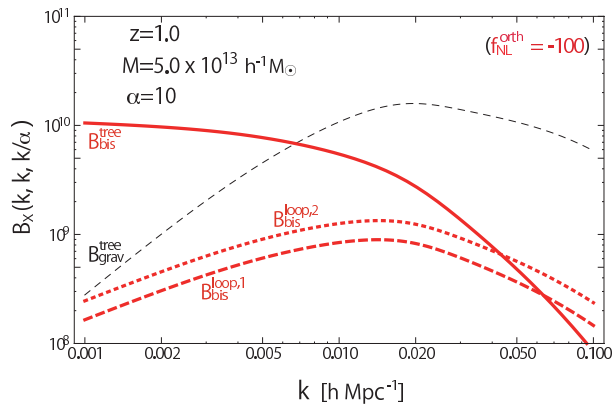}
   \end{center}
   \caption{$B_{\rm grav}^{\rm tree}$ (black dashed line), $B_{\rm bis}^{\rm tree}$ (red thick line), 
   $B_{\rm bis}^{{\rm loop},1}$ (red dashed thick line), and $B_{\rm bis}^{{\rm loop},2}$ (red dotted thick line)
   as a function of $k$  with fixing $\alpha=10$
   for orthogonal-type primordial non-Gaussianity. We take $f_{\rm NL}^{\rm orth} = -100$. }
   \label{fig:ort_loop.eps}
\end{figure}

Fig.~\ref{fig:ort_loop.eps} shows the scale-dependence of the halo bispectrum 
for the orthogonal type non-Gaussianity. In addition to the 
tree-level contributions, $B_{\rm grav}^{\rm tree}$ (black dashed line) and 
$B_{\rm bis}^{\rm tree}$ (red thick line), we also plot the one-loop contributions,  
$B_{\rm bis}^{{\rm loop},1}$ (red dashed thick line) and 
$B_{\rm bis}^{{\rm loop},2}$ (red dotted thick line) as functions of $k$, 
fixing $\alpha$ to $10$ and taking $f_{\rm NL}^{\rm orth} = -100$. 
Here, we do not show the contribution $B_{\rm bis}^{{\rm loop},3}$, because it 
turns out to be very small. As in the case of equilateral-type 
non-Gaussianity,  the one-loop contributions are much suppressed, and 
tree-level contributions $B_{\rm bis}^{\rm tree}$ becomes dominant on 
very large-scales.

\subsection{Comparison of total halo bispectrum between three types of 
primordial non-Gaussianity}

\begin{figure}[htbp]
 \begin{center}
  \includegraphics[width=150mm]{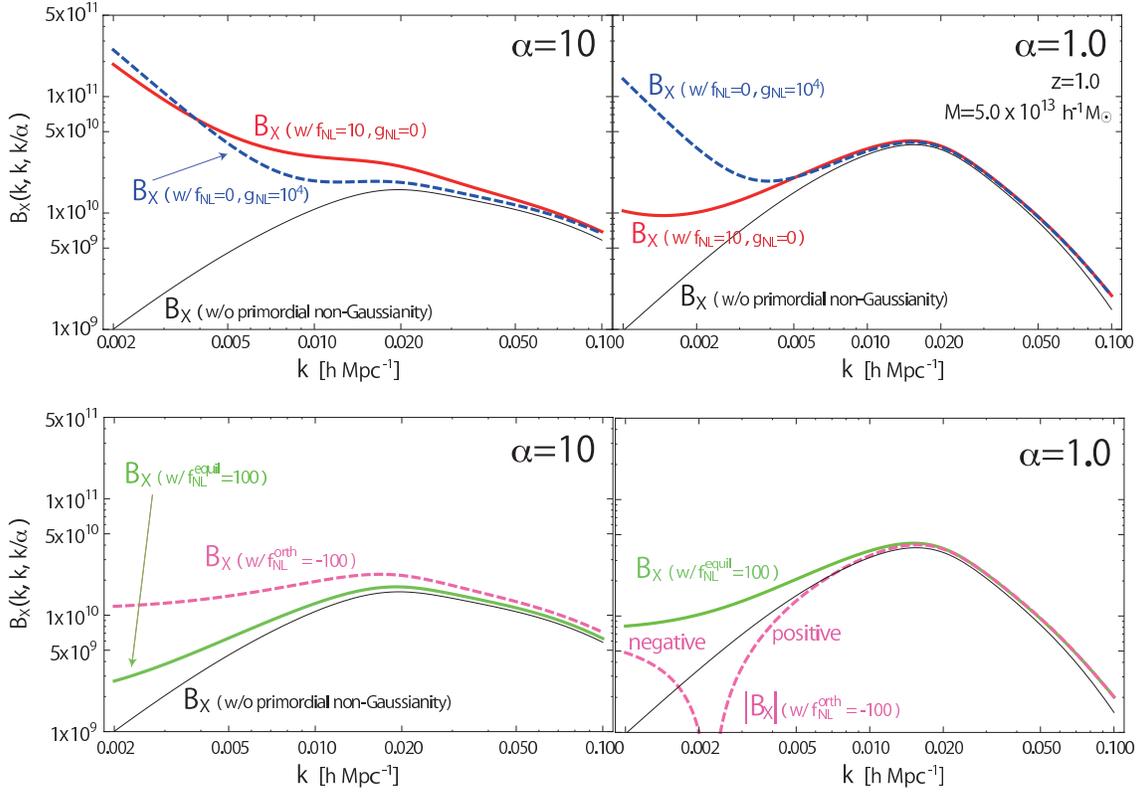}
   \end{center}
   \caption{Upper panels: total halo bispectrum, $B_X(k,k,k/\alpha)$, for local-type  with $\alpha=10$ (left) and $\alpha=1.0$ (right). 
   Black thin line is for the case without any primordial non-Gaussianity, red thick line is for the case with $f_{\rm NL} = 10$,
   $\tau_{\rm NL} = 36 f_{\rm NL}^2 / 25 $ and $g_{\rm NL} = 0$,
   and blue thick dashed line is for the case with $f_{\rm NL} = 0$ and $g_{\rm NL} = 10^4$. Lower panels:
   total halo bispectrum, $B_X(k,k,k/\alpha)$, for equilateral- (green thick line) and orthogonal-types (magenta thick dashed line) 
   with $\alpha=10$ (left) and $\alpha=1.0$ (right). We fix the non-linearity parameters to $f_{\rm NL}^{\rm equil} = 100$
   and $f_{\rm NL}^{\rm orth} = - 100$.
   In all panels, we fix the redshift and the mass of halos to $z=1.0$ and $M = 5 \times 10^{13} h^{-1} M_\odot$. }
   \label{fig:bis_total.eps}
\end{figure}

Finally, let us compare the overall trend of the total halo bispectrum at the 
one-loop order between three-types of primordial non-Gaussianity.  
Fig.~\ref{fig:bis_total.eps} shows the total halo bispectrum, 
$B_X(k,k,k/\alpha)$, as function of wavenumber, fixing $\alpha$ to 
$10$ (left) and $1.0$ (right). While upper panels present the results for 
the local-type non-Gaussianity with two different parameter set, i.e., 
$(f_{\rm NL},\,g_{\rm NL})=(10,\,0)$ and $(0,\,10^4)$, bottom panels 
plot the halo bispectra for the equilateral- (green) and orthogonal- types (magenta dashed). Here, we assume that the consistency relation, 
$\tau_{\rm NL} = 36 f_{\rm NL}^2 / 25 $, strictly holds for local-type 
non-Gaussianity. 
In each panel, black solid line indicates the result in the absence 
of the primordial non-Gaussianity, and considers only the contributions from 
the gravitational non-linearity.

For the non-linearity parameters consistent with current observations, 
the local-type non-Gaussianity gives the largest signal of the halo bispectrum 
in the squeezed case at large-scale. A notable point is that even with 
$\alpha=1.0$, a strong enhancement of the amplitude of bispectrum 
can be observed,  especially in the case of 
$(f_{\rm NL},\,g_{\rm NL})=(0,\,10^4)$. The scale-dependence 
of the halo bispectrum also appears in the cases of the equilateral- and orthogonal-type non-Gaussianities (see bottom panel). 
Although the effect is rather moderate, 
at large-scales, the contribution from primordial non-Gaussianity can 
exceed that from the gravitational non-linearity. Further, as we have seen 
in Sec.~\ref{sec:orthogonal}, additionally interesting 
feature can be seen for the orthogonal-type   
non-Gaussianity. With a 
negative non-linearity parameter $f_{\rm NL}^{\rm orth}<0$, 
the sign of the halo bispectrum eventually flips at large scales for the configuration with $\alpha\sim1$. This would be unique and important feature to 
distinguish the non-Gaussian signal from other types of non-Gaussianity. 
%

\section{Impact of higher-order contributions}
\label{sec:higherorder}

In the formalism of iPT, the contribution from the primordial 
non-Gaussianity on the halo bispectrum can be efficiently 
estimated from 
the diagrams including the linear (primordial) polyspectra. 
Increasing the order of perturbations, we can systematically calculate
the contributions from higher-order polyspectra. 
Naively, we expect that 
higher-loop contributions are generally suppressed on very large scales.  
As we saw in Sec.~\ref{sec:result_local}, however, 
one-loop contributions from the local-type non-Gaussianity 
are found to be non-negligible on large scales, 
and can eventually dominate the 
tree-level contributions. This partly implies that the perturbative 
treatment may not work well in the case of local-type non-Gaussianity, 
and the contribution from the two-loop order would be also dominant. 
In this section, focusing on the local-type non-Gaussianity, 
we study the impact of two-loop contribution on the halo bispectrum.

\subsection{Two-loop contributions from the primordial trispectrum}

As we found in Sec.~\ref{sec:result_local}, 
non-negligible contributions coming from the one-loop corrections
are described by the {\it un-decomposable} diagrams, which 
can not be simply constructed from tree-level diagrams by 
adding a primordial power spectrum (shown in Fig. \ref{fig:diagram_bis.eps}). 
In this respect, at two-loop order, 
possible non-negligible contributions may come from 
the diagrams linearly proportional to the primordial trispectrum or 
quadratically proportional to the primordial bispectrum. The other  
contributions linearly proportional to 
the primordial bispectrum would be certainly suppressed, since they 
are described as decomposable diagrams. 
\begin{figure}[htbp]
\begin{center}
 \includegraphics{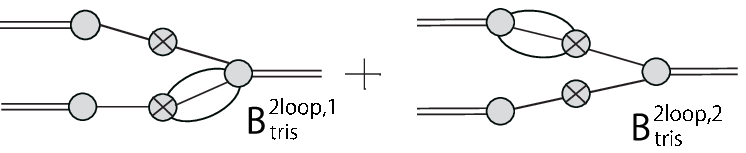}
\end{center}
\begin{center}
\includegraphics{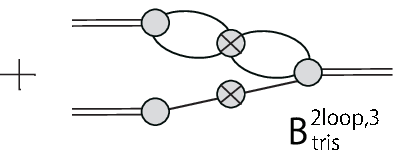}
\end{center}
\caption{Two loop diagrams linearly proportional to the primordial
trispectrum.}
\label{fig:lineargtau.eps}
\end{figure}

Let us first consider the two-loop un-decomposable contributions linearly proportional to the primordial trispectrum. The 
corresponding diagrams are shown in  Fig.~\ref{fig:lineargtau.eps}, each of 
which are analytically expressed as
\begin{eqnarray}
B_{\rm tris}^{2{\rm loop},1} &=&
{1 \over 6} \Biggl[
\Gamma_X^{(1)}(\bk_1) \Gamma_X^{(1)}(\bk_2) P_\rL(k_1) \cr\cr
&& \quad \times
\int {d^3 p_1 d^3 p_2 \over (2 \pi)^6}
\Gamma_X^{(4)} (-\bk_1, \bp_1, \bp_2, -\bk_2-\bp_1-\bp_2) T_\rL(\bk_2, \bp_1, \bp_2, -\bk_2 - \bp_1 - \bp_2) + 5~{\rm perms.} \Biggr], \cr\cr
B_{\rm tris}^{2{\rm loop},2} &=&
{1 \over 6}\Biggl[
\Gamma_X^{(1)}(\bk_1) \Gamma_X^{(2)}(-\bk_1, - \bk_3)P_\rL(k_1)
\cr\cr
&& \quad\times
\int {d^3 p_1 d^3 p_2 \over (2\pi)^6}
\Gamma_X^{(3)} (\bp_1, \bp_2, \bk_3 - \bp_1 -\bp_2)
T_\rL(-\bk_3, \bp_1, \bp_2, \bk_3 - \bp_1 -\bp_2) + 5~{\rm perms.} \Biggr],  \cr\cr
B_{\rm tris}^{2{\rm loop},3} &=&
{1 \over 4} \Biggl[
\Gamma_X^{(1)}(\bk_1) P_\rL(k_1)
\int {d^3 p_1 d^3 p_2 \over (2 \pi)^6}
\Gamma_X^{(2)}(\bp_1, \bk_2 - \bp_1) \cr\cr
&& \qquad\qquad\qquad \times
\Gamma_X^{(3)}(\bk_1, \bp_2, - \bk_2 - \bp_2)
 T_\rL(\bp_1, \bk_2 - \bp_1, \bp_2, -\bk_2 - \bp_2) + 5~{\rm perms.}
\Biggr].
\end{eqnarray}
In a manner similar to what we did in Sec.~\ref{sec:result_local}, 
we take the large-scale limit, $k_i \to 0$. Then, we have
\begin{eqnarray}
B_{\rm tris}^{2{\rm loop},1} &\approx&
\left(3 g_{\rm NL} + {25 \over 9}\tau_{\rm NL}\right)
\Gamma_X^{(1)}(\bk_1)\Gamma_X^{(1)}(\bk_2) P_\rL(k_1){P_\rL(k_2) \over {\cal M}(k_2)} \int {d^3 p_1 d^3 p_2 \over (2\pi)^6}
\Bigl[
-{\bk_1 \cdot \bk_3 \over k_1^2}c_3^\rL (\bp_1, \bp_2, - \bp_1-\bp_2) \cr\cr
&& \qquad\qquad\quad
+c_4^\rL(-\bk_1, \bp_1, \bp_2, -\bp_1 - \bp_2) \Bigr] {\cal M}(p_1){\cal M}(p_2){\cal M}(|\bp_1 + \bp_2|) P_\Phi(p_1)P_\Phi(p_2) + 5~{\rm perms.}
, \cr\cr
B_{\rm tris}^{2{\rm loop},2} &\approx&
\left(3 g_{\rm NL} + {25 \over 9}\tau_{\rm NL}\right)
\Gamma_X^{(1)}(\bk_1)\Gamma_X^{(2)}(-\bk_1, -\bk_3) P_\rL(k_1){P_\rL(k_3) \over {\cal M}(k_3)}
\cr\cr
&& \qquad \times
 \int {d^3 p_1 d^3 p_2 \over (2\pi)^6}c_3^\rL (\bp_1, \bp_2, - \bp_1-\bp_2){\cal M}(p_1){\cal M}(p_2){\cal M}(|\bp_1 + \bp_2|) P_\Phi(p_1)P_\Phi(p_2) + 5~{\rm perms.}, \cr\cr
B_{\rm tris}^{2{\rm loop},3} &\approx&
3 g_{\rm NL} \Gamma_X^{(1)}(\bk_1)P_\rL(k_1)
\Biggl\{ \int {d^3 p_1 \over (2\pi)^3} c_2^\rL(\bp_1, -\bp_1)P_\rL(p_1) \cr\cr
&&\qquad\qquad   \times \int {d^3 p_2 \over (2\pi)^3 }
\left[   -{\bk_1 \cdot \bk_3 \over k_1^2}c_2^\rL (\bp_2, -\bp_2) +c_3^\rL(-\bk_1, \bp_2,  - \bp_2)   \right] P_\rL(p_2) P_\Phi(p_2) \cr\cr
&& + \int {d^3 p_1 \over (2\pi)^3} c_2^\rL(\bp_1, -\bp_1)P_\rL(p_1)P_\Phi(p_1) \cr\cr
&& \qquad\qquad \times
\int {d^3 p_2 \over (2\pi)^3 }
\left[   -{\bk_1 \cdot \bk_3 \over k_1^2}c_2^\rL (\bp_2, -\bp_2) +c_3^\rL(-\bk_1, \bp_2,  - \bp_2)   \right] P_\rL(p_2)
\Biggr\} + 5~{\rm perms.} \cr\cr
&& + {25 \over 9}\tau_{\rm NL}\Gamma_X^{(1)}(\bk_1) P_\rL(k_1){P_\rL(k_2) \over {\cal M}(k_2)^2}
\int {d^3 p_1 \over (2\pi)^3} c_2^\rL(\bp_1, -\bp_1)P_\rL(p_1) \cr\cr
&& \qquad\qquad \times
 \int {d^3 p_2 \over (2\pi)^3 }
\left[   -{\bk_1 \cdot \bk_3 \over k_1^2}c_2^\rL (\bp_2, -\bp_2) +c_3^\rL(-\bk_1, \bp_2,  - \bp_2)   \right] P_\rL(p_2) + 5~{\rm perms.}.
\label{eq:Bgt2loop2}
\end{eqnarray}
Here, 
we used the fact that 
\begin{eqnarray}
\Gamma^{(4)}_X (-\bk_1, \bp_1,\bp_2, -\bk_2 -  \bp_1- \bp_2) \approx -{\bk_1 \cdot \bk_3 \over k_1^2}c_3^\rL (\bp_1, \bp_2, - \bp_1-\bp_2)
+c_4^\rL(-\bk_1, \bp_1, \bp_2, -\bp_1 - \bp_2) .
\end{eqnarray}
From the above expressions, we find 
\begin{eqnarray}
B_{\rm tris}^{2{\rm loop},1} \propto k^0 \alpha^1,~ B_{\rm tris}^{2{\rm loop},2} \propto k^0 \alpha^1,
~ B_{\rm tris}^{2{\rm loop},3} \propto k^{-2} \alpha^3~.
\label{eq:asympttwotri}
\end{eqnarray}
\begin{figure}[htbp]
 \begin{center}
  \includegraphics[width=80mm]{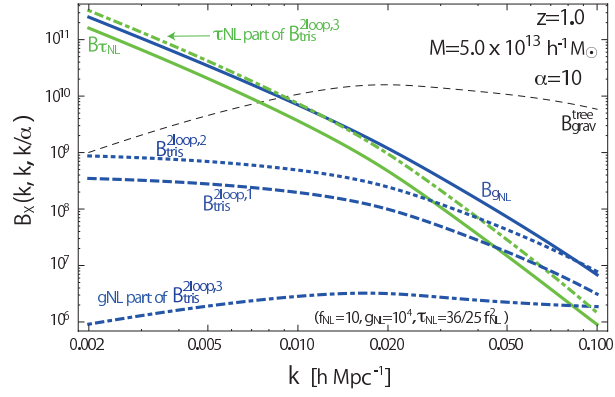}
   \end{center}
   \caption{ Two-loop contributions from the primordial trispectrum are plotted, which are
      $B_{\rm tris}^{2{\rm loop},1}$ (blue dashed thick line), $B_{\rm tris}^{2{\rm loop},2}$ (blue dotted thick line),
      $g_{\rm NL}$ contribution in
   $B_{\rm tris}^{2{\rm loop},3}$ (blue dot-dashed thick line), and $\tau_{\rm NL}$ contribution in $B_{\rm tris}^{2{\rm loop},3}$ (green dot-dashed thick line),  as functions of $k$.
   We also plot $B_{\rm grav}^{\rm tree}$ (black dashed line), $B_{g_{\rm NL}}$ (blue thick line), and $B_{\tau_{\rm NL}}$ (green thick line)
   as references. We take $g_{\rm NL} = 10^4$ and $\tau_{\rm NL} = (36/25)10^2$ and fix the redshift, the mass of halos and the squeezing parameter
   to $z=1.0$, $M = 5 \times 10^{13} h^{-1} M_\odot$, and $\alpha = 10$. }
   \label{fig:tri_two_loop.eps}
\end{figure}
In Fig.~\ref{fig:tri_two_loop.eps}, the three two-loop contributions, 
$B_{\rm tris}^{2{\rm loop},1}$ (blue dashed thick line), $B_{\rm tris}^{2{\rm loop},2}$ (blue dotted thick line), and $B_{\rm tris}^{2{\rm loop},3}$, are plotted 
 as functions of $k$, fixing $\alpha$ to $10$. 
For the contribution, $B_{\rm tris}^{2{\rm loop},3}$, we divide it into two pieces, 
and separately show 
the contributions proportional to $g_{\rm NL}$ (blue dot-dashed thick line) 
and $\tau_{\rm NL}$ (green dot-dashed thick line). 
Then, all the un-decomposable two-loop contributions 
proportional to $g_{\rm NL}$ turn out to be sub-dominant on large scales,  
while the term proportional to $\tau_{\rm NL}$ (labeled as 
$\tau_{\rm NL}$ part of $B_{\rm tris}^{2{\rm loop},3}$) gives a 
significantly large contribution, which can dominate over
the one-loop contribution, $B_{\tau_{\rm NL}}$. 
This implies that 
we need to develop at least the two-loop order calculations if one wants to 
evaluate the contribution of the halo bispectrum 
linearly proportional to the non-linearity parameter $\tau_{\rm NL}$. 
As for the case of $g_{\rm NL}$, the one-loop order calculation seems sufficient 
on large scales. 

\subsection{Other two-loop contributions}

\begin{figure}[htbp]
\begin{center}
 \includegraphics{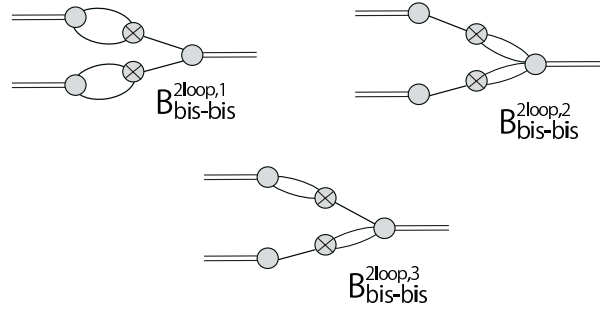}
\end{center}
\caption{Undecomposable diagrams which are quadratically proportional to the primordial bispectrum. }
\label{fig:diagram_ff.eps}
\end{figure}

Consider next the un-decomposable two-loop contributions quadratically 
proportional to the primordial bispectrum, which are 
diagrammatically shown in Fig.~\ref{fig:diagram_ff.eps}.
For local-type non-Gaussianity, the inequality,  
$\tau_{\rm NL} \geq 36 f_{\rm NL}^2 / 25$, generally holds. In the case of equality which we consider here, this implies that the diagrams of the $\mathcal{O}(f_{\rm NL}^2)$ order are expected to give a significant contribution comparable to the $\tau_{\rm NL}$ contribution, and potentially become dominant on large scales. 
In the large-scale limit, we approximately have
\begin{eqnarray}
B_{\rm bis-bis}^{2{\rm loop},1} &=& {1 \over 4} \Gamma_X^{(2)}(\bk_1,\bk_2) 
\int {d^3 p_1 d^3 p_2 \over (2 \pi)^6} \Gamma_X^{(2)}(\bp_1, \bk_1 - \bp_1) \Gamma_X^{(2)}(\bp_2, \bk_2 - \bp_2) \cr\cr
&& \qquad\qquad\qquad \times
B_\rL(-\bk_1, \bp_1, \bk_1 - \bp_1) B_\rL(-\bk_2, \bp_2, \bk_2 - \bp_2) + 2~{\rm perms.} \cr\cr
&\approx&
4f_{\rm NL}^2 \Gamma_X^{(2)}(\bk_1,\bk_2)
{P_\rL(k_1)P_\rL(k_2) \over {\cal M}(k_1){\cal M}(k_2)} \left[ \int {d^3 p \over (2 \pi)^3} c_2^\rL(\bp, - \bp)P_\rL(p)\right]^2
+ 2~{\rm perms.}, \cr\cr
B_{\rm bis-bis}^{2{\rm loop},2} 
&\approx&
4f_{\rm NL}^2 \Gamma_X^{(1)}(\bk_1)\Gamma_X^{(1)}(\bk_2) \int {d^3p_1d^3p_2  \over (2\pi)^6}
c_4^\rL(\bp_1, - \bp_1,\bp_2, - \bp_2) 
P_\rL(p_1)P_\rL(p_2)  {P_\rL(k_1) P_\rL(k_2) \over {\cal M}(k_1){\cal M}(k_2)}  + 2~{\rm perms.},\cr\cr
B_{\rm bis-bis}^{2{\rm loop},3} 
&\approx&
4f_{\rm NL}^2 \Gamma_X^{(1)}(\bk_1)  {P_\rL(k_1) P_\rL(k_2) \over {\cal M}(k_1){\cal M}(k_2)} \int {d^3p_1  \over (2\pi)^3}
c_2^\rL(\bp_1, - \bp_1) 
P_\rL(p_1)\cr\cr
&& \times \int  {d^3 p_2 \over (2\pi)^3} \left[ - {\bk_2 \cdot \bk_3 \over k_2^2} c_2^\rL(-\bp_2, \bp_2)
+ c_3^\rL(-\bk_2, - \bp_2, \bk_1 + \bp_2) \right] P_\rL (p_2)  + 5~{\rm perms.}~.
\end{eqnarray}
Then, the asymptotic behavior of these contributions becomes
\begin{eqnarray}
B_{\rm bis-bis}^{2{\rm loop},1} \propto k^{-2} \alpha~, ~B_{\rm bis-bis}^{2{\rm loop},2} \propto k^{-2} \alpha~,~
B_{\rm bis-bis}^{2{\rm loop},3} \propto k^{-2} \alpha~.
\label{eq:asymptbisbis}
\end{eqnarray}
%
%
\begin{figure}[htbp]
 \begin{center}
  \includegraphics[width=80mm]{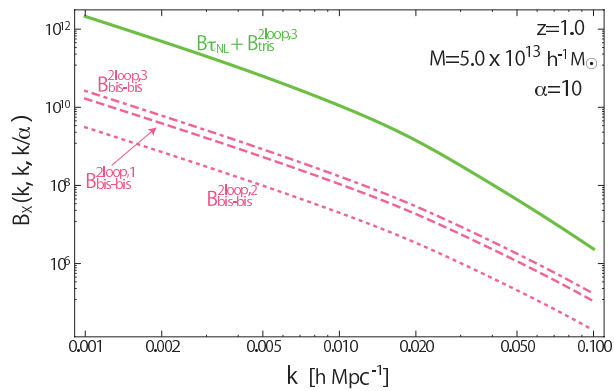}
   \end{center}
   \caption{$B_{\tau_{\rm NL}}(k,k,k/\alpha)+B_{\rm tris}^{2{\rm loop},3}(k,k,k/\alpha)$ (green thick line),
   $B_{\rm bis-bis}^{2{\rm loop},1}$ (magenta dashed line), $B_{\rm bis-bis}^{2{\rm loop},2}$ (magenta dotted line),
   and $B_{\rm bis-bis}^{2{\rm loop},3}$ (magenta dot-dashed line)
    as a function of $k$  with fixing $\alpha=10$. We take $f_{\rm NL}^{\rm local} = 10$ and $\tau_{\rm NL}=36 f_{\rm NL}^2/25$.}
   \label{fig:fNLfNL.eps}
\end{figure}
In Fig.~\ref{fig:fNLfNL.eps}, the two-loop contributions given above are 
plotted as functions of $k$, fixing $\alpha$ to $10$;  
$B_{\rm bis-bis}^{2{\rm loop},1}$ (magenta dashed), $B_{\rm bis-bis}^{2{\rm loop},2}$ (magenta dotted), and $B_{\rm bis-bis}^{2{\rm loop},3}$ (magenta dot-dashed). For comparison, we also plot the one- and two-loop contributions linearly 
proportional to $\tau_{\rm NL}$, i.e., $B_{\tau_{\rm NL}}(k,k,k/\alpha)+B_{\rm tris}^{2{\rm loop},3}(k,k,k/\alpha)$ (green thick), assuming 
$f_{\rm NL}^{\rm local} = 10$ and $\tau_{\rm NL}=36 f_{\rm NL}^2/25$. 
Then, all the two-loop contributions quadratically proportional to the 
primordial bispectrum can become sub-dominant, and are well below 
the $\tau_{\rm NL}$-contribution. The asymptotic behaviors given in 
Eqs. (\ref{eq:asymptotic}), (\ref{eq:asympttwotri}), and 
(\ref{eq:asymptbisbis}) indicate that all the contributions shown in 
Fig.~\ref{fig:fNLfNL.eps} scale as $k^{-2}$, but the terms 
$B_{\tau_{\rm NL}}$ and $B_{\rm tris}^{2{\rm loop},3}$ have a larger power of $\alpha$, and are proportional to $\alpha^3$, which results in a significantly larger 
amplitude than that of  
$B_{\rm bis-bis}^{2{\rm loop}}$ by two orders of magnitude.

To sum up, it is not always the case that the higher loop contributions are suppressed compared with the lower loop contributions.  
In particular, the un-decomposable loop diagrams in the context of iPT would 
produce dominant contributions in the large-scale halo bispectrum, 
and hence we need to take into account all of the un-decomposable diagrams 
in order to precisely evaluate the effect of 
primordial non-Gaussianity on the halo bispectrum. 
In doing this, the diagrammatic approach based on the iPT helps 
us to systematically collect these dominant contributions.

\section{Discussion}
\label{sec:discuss}

In this section, we discuss several points which have not been 
yet clarified in previous sections. First, in Sec.~\ref{subsec:halo_mass_z},  
we examine the dependence of redshift and 
halo mass on the halo bispectrum in the case of local-type non-Gaussianity. 
In Sec.~\ref{subsec:asymptotic_scale-dependence}, to infer the 
asymptotic behavior of the bispectrum, we give a simple estimate of 
the scale-dependent behaviors using the diagrammatic approach. Finally, in 
Sec.~\ref{subsec:comparison}, the results of our analysis 
are compared with those of recent other works.

\subsection{Dependence of redshift and mass of halos on the bispectrum }
\label{subsec:halo_mass_z}

So far, we have set the redshift and the mass scale of halos  to 
$z=1.0$ and $M = 5 \times 10^{13} h^{-1} M_\odot$ as representative values 
relevant for observations. 
Here, focusing on the local-type non-Gaussianity and based on the formulae 
for bispectrum at one-loop order, 
we discuss the redshift and the mass dependences of the halo bispectrum. 

\subsubsection{redshift}
\label{appsec:redshift}

\begin{figure}[htbp]
 \begin{center}
  \includegraphics[width=170mm]{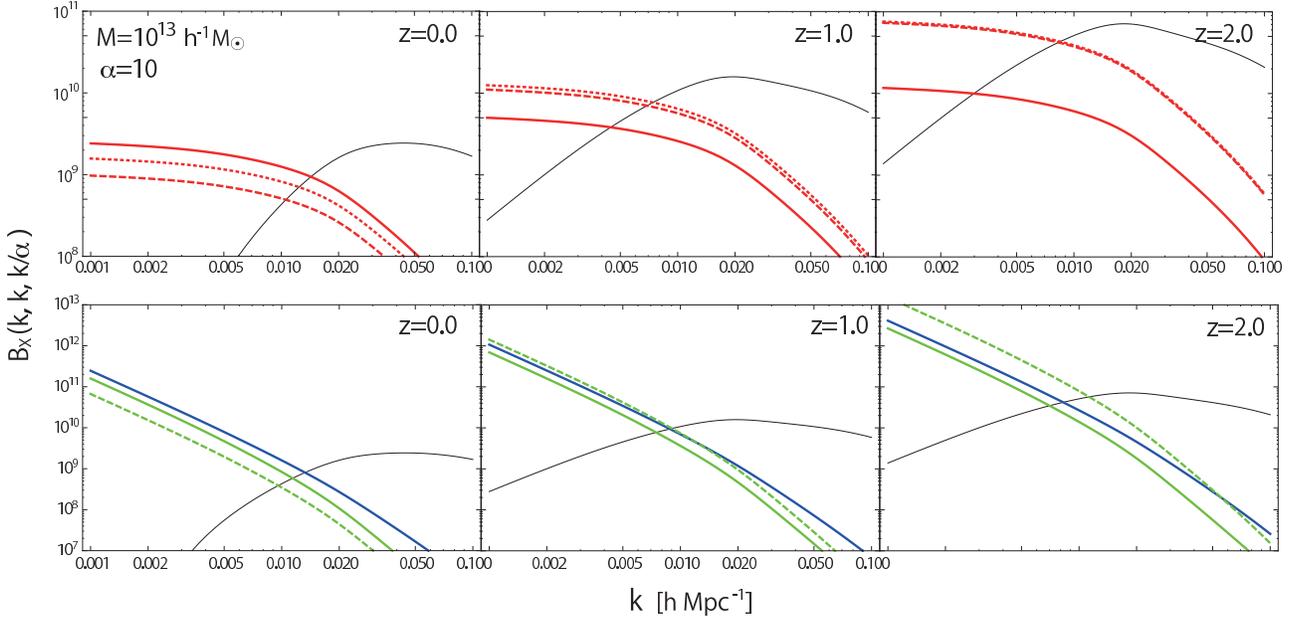}
   \end{center}
   \caption{Each contribution of the halo bispectrum, Top - $B_{\rm grav}$ (black), $B_{\rm bis}^{\rm tree}$ (red thick), $B_{\rm bis}^{{\rm loop},1}$ (red thick dashed),
   and $B_{\rm bis}^{{\rm loop},2}$ (red thick dotted), Bottom - $B_{\rm grav}$ (black), $B_{g_{\rm NL}}$ (blue thick), $B_{\tau_{\rm NL}}$ (green thick),
   and $B_{\rm tris}^{2{\rm loop},3}$ (green thick dashed) with changing the mass of halo as $ z=0.0 $ (left), $z=1.0$ (central), and the right for $z=2.0$ (right).}
   \label{fig:fgtnlz.eps}
\end{figure}

The redshift dependence of the halo bispectrum is shown in Fig. \ref{fig:fgtnlz.eps}.
In this figure, we plot each contribution of the halo bispectrum, $B_{\rm grav}$ (black), $B_{\rm bis}^{\rm tree}$ (red thick), $B_{\rm bis}^{{\rm loop},1}$ (red thick dashed),
   and $B_{\rm bis}^{{\rm loop},2}$ (red thick dotted) in top panels, and  
   $B_{\rm grav}$ (black), $B_{g_{\rm NL}}$ (blue thick), $B_{\tau_{\rm NL}}$ (green thick),
   and $B_{\rm tris}^{2{\rm loop},3}$ (green thick dashed) in bottom panels.
   From left to right, we change
   the redshift as $ z=0.0$, $1.0$, and $2.0$.
For these redshifts, we have checked that the other contributions are also suppressed.
From this figure, we find that the scale-dependence of the halo bispectrum does not change with the redshift,
but the most dominant contribution changes. 
The redshift parameter has an affect on not only the evolution of the density perturbations but also $b_n^\rL$.
Basically, the density perturbations grow as the redshift parameter decreases.
On the other hand, once the mass is fixed, the halo at the higher redshift becomes more highly biased object and it means larger value of $b_n^\rL$.
As shown in Fig. \ref{fig:fgtnlz.eps}, the halo bispectrum 
becomes larger at the higher redshift and hence we find that the halo bispectrum seems to 
have the stronger dependence of the redshift parameter through $b_n^\rL$ than the growth function $D(z)$.
As for the most dominant contribution, we find for the more highly biased object the higher order
loop contribution becomes important.
Basically, the higher order loop contribution depends on the higher order scale-independent
Lagrangian bias parameter $b_n^\rL$ and such higher order $b_n^\rL$ tends to become larger
for the more highly biased object.  
Hence, the higher order loop contribution becomes important for the higher redshift, as can be seen in Fig.  \ref{fig:fgtnlz.eps}.

\subsubsection{mass of the halos}

\begin{figure}[htbp]
 \begin{center}
  \includegraphics[width=170mm]{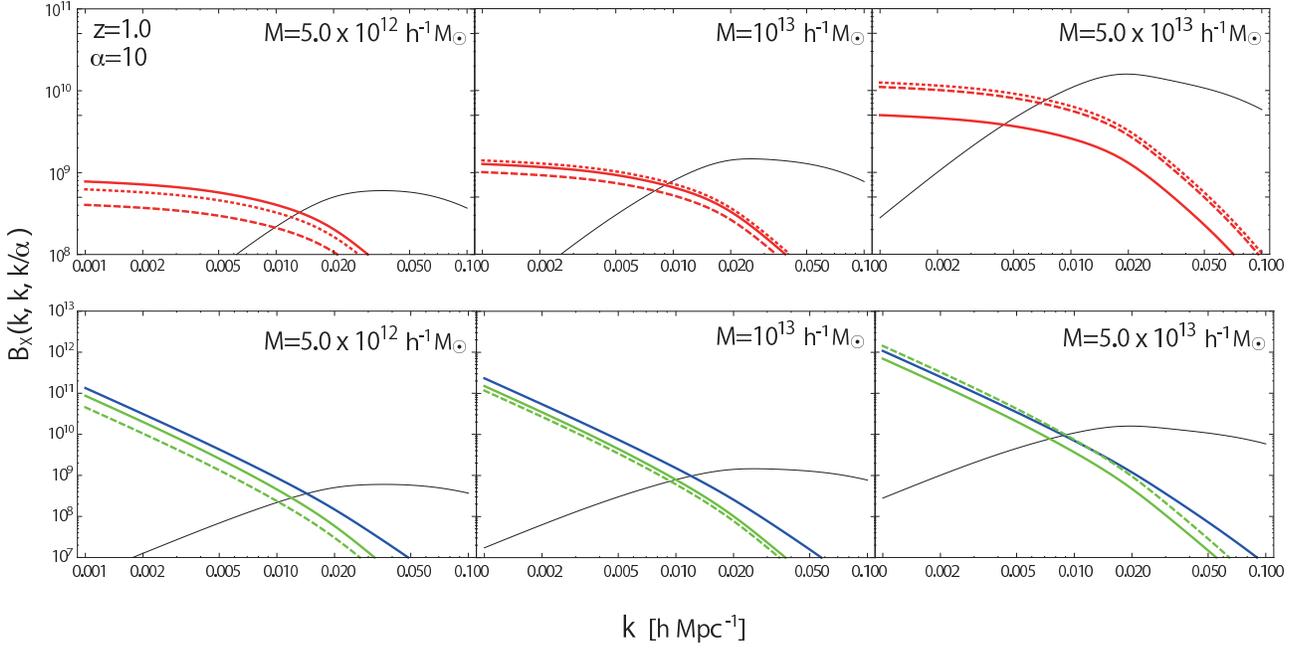}
   \end{center}
   \caption{Each contribution of the halo bispectrum, Top - $B_{\rm grav}$ (black), $B_{\rm bis}^{\rm tree}$ (red thick), $B_{\rm bis}^{{\rm loop},1}$ (red thick dashed),
   and $B_{\rm bis}^{{\rm loop},2}$ (red thick dotted), Bottom - $B_{\rm grav}$ (black), $B_{g_{\rm NL}}$ (blue thick), $B_{\tau_{\rm NL}}$ (green thick),
   and $B_{\rm tris}^{2{\rm loop},3}$ (green thick dashed) with changing the mass of halo as $ M = 5.0 \times 10^{12} h^{-1} M_\odot$ (left), $M=10^{13} h^{-1} M_\odot$ (central), and $M=5.0 \times
   10^{13} h^{-1} M_\odot$ (right).}
   \label{fig:fgtnlm.eps}
\end{figure}

The mass dependence of the halo bispectrum is shown in Fig. \ref{fig:fgtnlm.eps}.
In this figure, we plot each contribution of the halo bispectrum,  $B_{\rm grav}$ (black), $B_{\rm bis}^{\rm tree}$ (red thick), $B_{\rm bis}^{{\rm loop},1}$ (red thick dashed),
   and $B_{\rm bis}^{{\rm loop},2}$ (red thick dotted) in top panels, and  
   $B_{\rm grav}$ (black), $B_{g_{\rm NL}}$ (blue thick), $B_{\tau_{\rm NL}}$ (green thick),
   and $B_{\rm tris}^{2{\rm loop},3}$ (green thick dashed) in bottom panels.
   From left to right, we change
   the mass of halo as $ M = 5.0 \times 10^{12} h^{-1} M_\odot$, $10^{13} h^{-1} M_\odot$, and $5.0 \times
   10^{13} h^{-1} M_\odot$.
   As in the discussion about the redshift dependence,
   we have checked that the other contributions are suppressed, and
we also find that the scale-dependence of the halo bispectrum does not change with changing the mass of halo,
but the dominant contribution changes. 
The dependence of the mass of halos is reflected in the scale-independent Lagrangian bias parameter $b_n^\rL$,
and it becomes larger for the larger mass.
As we discussed before,
higher order $b_n^\rL$ tends to become larger for the larger mass
and hence the higher order loop contribution becomes more important for the larger mass of the halos, as can be seen
in Fig. \ref{fig:fgtnlm.eps}.

As the result, although which loop contribution dominates in the halo bispectrum on large scales
depends on the bias parameter $b_n^\rL$, that is, the redshift and the mass of observed halos,
we stress that it should be essential to consider both contributions of one and two-loops coming from the
primordial trispectrum.

\subsection{Asymptotic scale-dependence}
\label{subsec:asymptotic_scale-dependence} 

The asymptotic scale-dependence is important 
to find dominant contributions of the halo bispectrum 
on large scales. In the discussion above, starting  
with the analytic expression of each diagram, we took
the large-scale limit, and have finally derived the formulae for asymptotic 
scale-dependence. Here, we will argue that 
using the diagrammatic approach, these formulae can be recovered 
more easily and systematically.

First consider the simple case, and 
look at the contribution, $B_{\rm grav}^{\rm tree}$, shown in 
Fig.~\ref{fig:diagram_bis.eps}.  This diagram  
can be graphically divided into two pieces, consisting of 
linear power spectrum of the biased object, $P_X^{\rm lin}$. Thus, 
we may write it as $B_{\rm grav}^{\rm tree} \propto P_X^{\rm lin} \times P_X^{\rm lin}$.
For scale-invariant primordial fluctuations, 
the linear power spectrum $P_X^{\rm lin}$ is proportional to $k$ 
on large scales, and 
we immediately obtain $B_{\rm grav}^{\rm tree} (k_1,k_2,k_3)  \propto 
P_X^{\rm lin}(k_1) \times P_X^{\rm lin}(k_2) + 2~{\rm perms.} \propto k^2\alpha^0$, 
as a dominant contribution on large scales ($k \to 0$) 
in the squeezed limit ($\alpha \to \infty $).

Similarly, we can also divide the contribution of the diagrams 
$B_{\rm bis}^{{\rm loop},1}$ and $B_{\rm bis}^{{\rm loop},2}$ in 
Fig.~\ref{fig:diagram_bis.eps} into two pieces. 
In this case, a part of the diagram is $P_X^{\rm lin}$, 
but another piece corresponds to the one-loop power spectrum 
of the biased object, induced by the primordial bispectrum, $P_X^{\rm loop}$. 
It is known in the literature that  
the local-type primordial bispectrum
generates a strong scale-dependence in the bias parameter, which scales 
as $k^{-2}$.  In iPT, the main contribution of this indeed comes from 
the one-loop power spectrum,  and 
we have $P_X^{\rm loop} \propto P_X^{\rm lin} / k^2 \propto k^{-1}$. 
Hence, we reproduce the asymptotic scale-dependence of 
$B_{\rm bis}^{{\rm loop},1}$ and $B_{\rm bis}^{{\rm loop},2}$, and we obtain
\begin{eqnarray} 
B_{\rm bis}^{{\rm loop},1(2)} (k_1,k_2,k_3) \propto P_X^{\rm lin}(k_1) \times P_X^{\rm loop}(k_2) + 2~{\rm perms.}
\propto k^{0} \alpha, 
\nonumber
\end{eqnarray}
as a dominant contribution on large scales in the squeezed limit. 
This systematic approach based on the diagrammatic picture can also apply to 
the other contributions, and 
we easily find their asymptotic scale-dependences. 

\subsection{Comparison with other works}
\label{subsec:comparison}

Finally, let us compare the results of our analysis with those of 
the recent works. In Ref.~\cite{Jeong:2009vd}, adopting 
the Taylor-expansion with local bias ansatz, 
the authors derived the formula for the halo bispectrum based on the 
Matarrese-Lucchin-Bonometto (MLB)
formalism \cite{Matarrese:1986et}.  MLB formalism is nothing but the 
peak formalism, and we confirmed that taking the high-peak limit, 
our formula at the one-loop order 
reproduces the result in Ref.~\cite{Jeong:2009vd}.  
In the present paper, we further investigated the two-loop contribution, and 
found a new dominant contribution from the primordial trispectrum,  
which has not been considered in Ref.~\cite{Jeong:2009vd}.

As another analytic approach, Ref.~\cite{Baldauf:2010vn} has 
presented a formula for the halo bispectrum based on 
the peak-background split picture. The authors particularly examined 
the case of the single-component non-Gaussianity characterized by the 
curvature perturbation, 
$\Phi = \Phi_G + f_{\rm NL} ( \Phi_G^2 - \langle \Phi_G^2\rangle )
+ g_{\rm NL} \Phi_G^3$. 
Following Ref.~\cite{Matsubara:2012nc}, and 
using the cancellation properties of the highest-order 
parameters in Press-Schechter mass function, 
we have checked that our formula is consistent with Eq.~(5.1) in 
Ref.~\cite{Baldauf:2010vn} in the large-scale limit. 
Note that the bispectrum formula in Ref.~\cite{Baldauf:2010vn}
additionally includes contributions 
proportional to $f_{\rm NL}^3$ or $f_{\rm NL}^4$, 
for which we did not consider here. 
As shown in the figures in Ref.~\cite{Baldauf:2010vn}, however, 
these contributions can become 
significant only at very much large scales ($k \lesssim 0.002 h {\rm Mpc}^{-1}$) 
in the squeezed isosceles configuration ($\alpha = 10$).  
Ref.~\cite{Nishimichi:2009fs} investigated the halo 
bispectrum in cosmological $N$-body simulations, 
and found that the results of their simulations are  
rather consistent with those predicted by Ref.~\cite{Jeong:2009vd}, and 
the dominant contribution of bispectrum 
scales as $f_{\rm NL}^2$ in the squeezed limit. 
Hence, the contributions proportional to $f_{\rm NL}^3$ or $f_{\rm NL}^4$ 
might not be relevant for real observations.

Although we do not discuss in detail, 
we note here that these contributions 
can be also constructed in the formalism of the iPT. 
Such contributions 
come from the primordial higher-order poly-spectra as follows.
For the single-sourced case,  
the primordial higher-order spectra can be simply parameterized by
the non-linearity parameter $f_{\rm NL}$. 
For example, the primordial 5- and 6-point spectra respectively have the dependence of $f_{\rm NL}^3$ and $f_{\rm NL}^4$. 
Including such primordial higher-order poly-spectra and considering corresponding undecomposable diagrams, 
we can recover the same contributions as found 
in Ref.~\cite{Baldauf:2010vn}. 
We also find a crude relationship between the contributions from $\tau_{\rm NL}$ and 
that from $f_{\rm NL}^3$ and $f_{\rm NL}^4$ as
$ B_{f_{\rm NL}^3} \sim  B_{\tau_{\rm NL}} \times (f_{\rm NL} / {\cal M}(k)) $
and $ B_{f_{\rm NL}^4} \sim  B_{\tau_{\rm NL}} \times (f_{\rm NL} / {\cal M}(k))^2 $, 
where we respectively denote the contributions from $f_{\rm NL}^3$ and $f_{\rm NL}^4$
as $B_{f_{\rm NL}^3}$ and $B_{f_{\rm NL}^4}$.
From this relation, in order for $B_{f_{\rm NL}^3}$ and $B_{f_{\rm NL}^4}$
to dominate over $B_{\tau_{\rm NL}}$,
we need $k < 10^{-3} h {\rm Mpc}^{-1}$ for $f_{\rm NL} = 10$.
Thus, we stress here that for the accessible scale of observations 
and with the currently allowed values of non-linearity parameters, 
one- and two-loop contributions coming from the primordial trispectrum 
are important for the halo bispectrum.

\begin{figure}[htbp]
 \begin{center}
  \includegraphics[width=150mm]{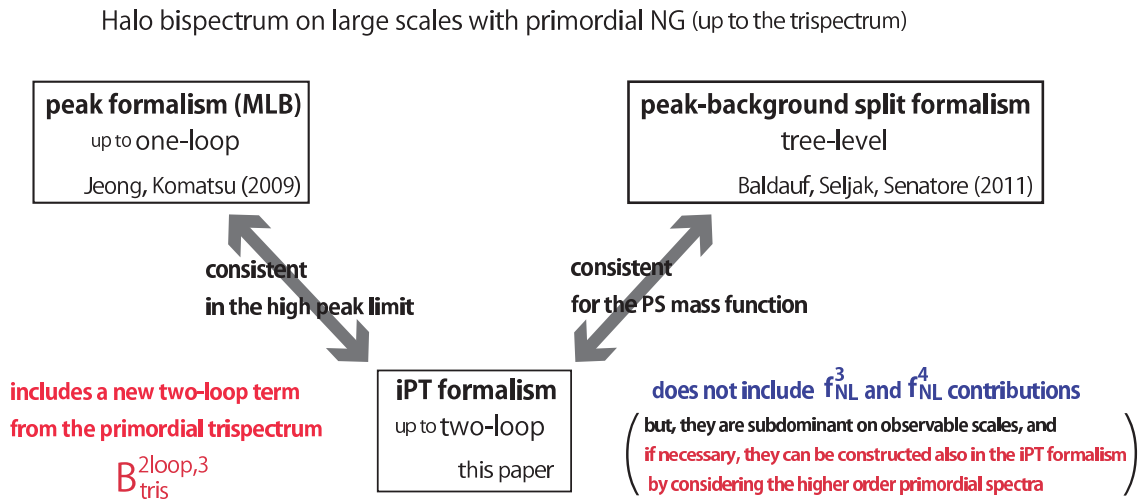}
   \end{center}
   \caption{Summary of the consistency between three analytic formalism for the halo bispectrum with the primordial non-Gaussianity.}
   \label{fig:table1.eps}
\end{figure}

In Fig.~\ref{fig:table1.eps}, we summarize the consistency between 
the three analytic formalism for the halo bispectrum in the presence of 
the primordial non-Gaussianity. 

\section{Summary}
\label{sec:sum}

In this paper, on the basis of the iPT formalism,
we have systematically 
investigated the bispectrum of the biased objects induced by the primordial 
non-Gaussianity.
Basically, our formulae in the case of local-type non-Gaussianity 
are consistent with those of the previous works. Notable point in the present 
paper is that we studied the halo bispectrum in the case of 
not only the local-type, but also the 
equilateral- and orthogonal-type primordial non-Gaussianity. 
Further, in case of the local-type, we include the effect of the primordial 
trispectrum 
characterized by the two non-linearity parameters, $g_{\rm NL}$ and $\tau_{\rm NL}$.
For the local-type non-Gaussianity, the two-loop correction coming from the 
terms proportional to $\tau_{\rm NL}$, denoted by $B_{\rm tris}^{2{\rm loop},3}$, is found to produce a non-negligible 
contribution, whose amplitude can become comparable to the 
dominant one-loop corrections proportional to $f_{\rm NL}^2$, denoted by $B_{\tau_{\rm NL}}$. 
As discussed in the previous section, 
such non-negligible loop contributions are represented by
un-decomposable loop diagrams in the context of iPT, and   
taking account of such un-decomposable diagrams is rather crucial 
to precisely investigate the dominant contributions 
from the primordial non-Gaussianity. 
We stress here that the diagrammatic approach based on the iPT 
helps us to systematically collect these dominant contributions. 
The significance of higher-order contributions in the case of 
local-type non-Gaussianity is our important findings, and it 
implies that a much more great impact on the convergence of perturbative 
expansion is expected in the presence of higher-order primordial polyspectra.  
This should deserve further investigation, and we 
need a more clear and physical understanding of our results.

For the current observational upper-bound of the non-linearity parameters, 
the contributions from the primordial trispectrum would dominate 
the halo bispectrum on large scales. 
Taking the typical redshift and the mass of halos in surveys to be $z=1.0$
and $M=5 \times 10^{13} h^{-1} M_\odot$ respectively, 
the halo bispectrum with $g_{\rm NL} = 10^4$ 
is comparable to the one induced by 
$\tau_{\rm NL} = 36 f_{\rm NL}^2/25$ with $f_{\rm NL}=10$.
In Ref.~\cite{Sefusatti:2007ih}, 
the authors mentioned that the future galaxy surveys
would have a potential to detect $f_{\rm NL} \sim O(10)$, and 
this is expected to give $g_{\rm NL} < O(10^4)$. 
The current observational limit on $g_{\rm NL}$ is about $g_{\rm NL} < O(10^5)$
and hence the bispectrum of the biased object would be a powerful 
tool to obtain
tighter constraint on $g_{\rm NL}$.
For future idealistic wide-field surveys at $z=1$, in which we 
have a survey volume of $V=10 h^{-3} {\rm Gpc}^3$ and the observable 
number density of the halos is $\bar{n} = 10^{-4} h^3 {\rm Mpc}^{-3}$ with 
typical mass $M=5 \times 10^{13} h^{-1} M_\odot$,
we expect that the detectability is improved to $g_{\rm NL} = O(10^4)$.
It is therefore very interesting to pursue a precise forecast for $g_{\rm NL}$
from the halo/galaxy bispectrum data obtained by the future surveys.

For equilateral- and orthogonal-types of non-Gaussianity, 
we find that the 
one-loop corrections are all suppressed and the tree-level calculation 
is sufficient to investigate the dominant contribution of halo bispectrum. 
Following the result obtained by Ref.~\cite{Sefusatti:2007ih}, 
future galaxy surveys can have a potential to 
obtain a constraint on $f_{\rm NL}^{\rm equil}$ down to $f_{\rm NL}^{\rm equil}< O(50)$, 
and it is comparable to the constraint
obtained from the CMB observations. We expect that such future surveys also give a tight constraint on $f_{\rm NL}^{\rm orth}$.
In this paper, we did not consider the contribution from the trispectrum 
in the cases of equilateral- and orthogonal-types non-Gaussianity,  
because their scale dependence is strongly dependent of the model 
generating the primordial non-Gaussianity. Also, the exact form of 
primordial trispectrum 
is much more complicated compared to that of the local-type non-Gaussianity.
We leave the discussion on the contribution from the equilateral or 
orthogonal trispectrum to future work. 
Finally, while the recent cosmological $N$-body simulations suggest that the 
measured halo bispectrum on large scales is 
consistent with the previous analytic formulae 
Ref.~\cite{Nishimichi:2009fs,Sefusatti:2011gt,Baldauf:2010vn}, 
we believe that our formalism is capable of giving much more precise prediction 
which quantitatively reproduces the simulation results 
for various types of the primordial non-Gaussianity even on smaller scales. 
Along the line of this, the detailed comparison with $N$-body simulation is 
our important next subject.

\acknowledgments
This work was supported by the
Grant-in-Aid for JSPS Research under Grant No. 24-2775 (SY), Grant-in-Aid for Scientific Research (C), 24540267, 2012 (TM)
and also 24540257 (AT).
To complete this work, discussions during the YITP workshop YITP-T-13-06 on "The CMB and theory of the primordial universe" were useful, and we thank YITP.

\appendix

\section{The choice of the mass function}
\label{sec:appendix}

\begin{figure}[htbp]
 \begin{center}
  \includegraphics[width=170mm]{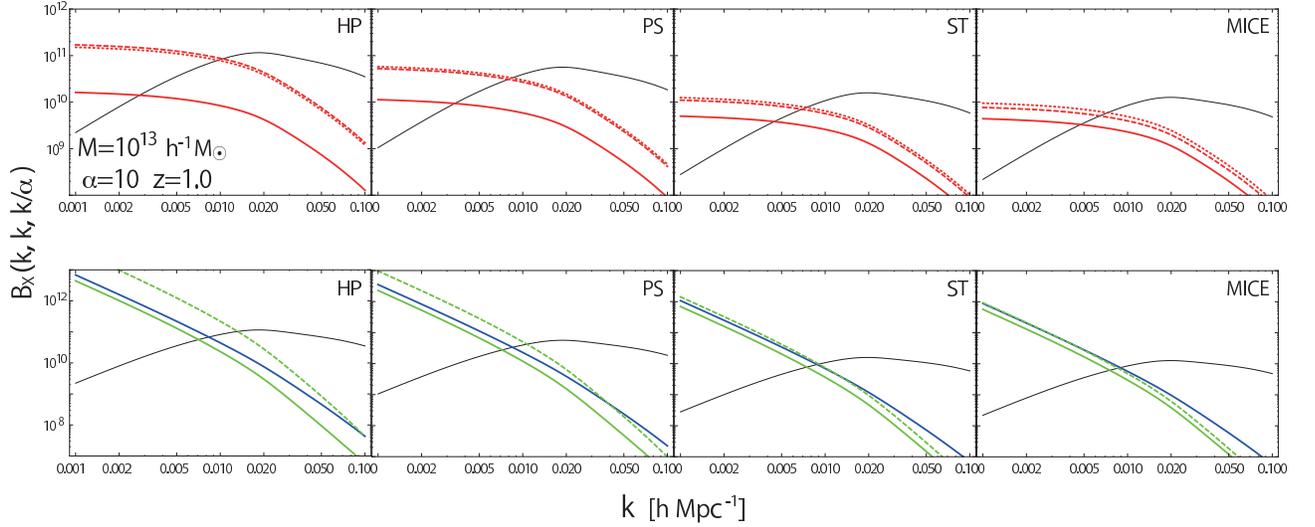}
   \end{center}
   \caption{Each contribution of the halo bispectrum, Top - $B_{\rm grav}$ (black), $B_{\rm bis}^{\rm tree}$ (red thick), $B_{\rm bis}^{{\rm loop},1}$ (red thick dashed),
   and $B_{\rm bis}^{{\rm loop},2}$ (red thick dotted), Bottom - $B_{\rm grav}$ (black), $B_{g_{\rm NL}}$ (blue thick), $B_{\tau_{\rm NL}}$ (green thick),
   and $B_{\rm tris}^{{\rm loop},3}$ (green thick dashed) with tthe scale-independent bias parameters given by Eq. (\ref{eq:HPMF}) (HP),
   and the mass function given by (\ref{eq:PSMF}) (PS), (\ref{eq:STMF}) (ST) and (\ref{eq:MICEMF}) (MICE) from left to right.}
   \label{fig:fgtnlmf.eps}
\end{figure}

Throughout this paper, 
we apply the Sheth-Tormen fitting formula as a mass function in the derivation of the scale-independent bias parameter $b_n^\rL$.
Let us discuss the dependence of the halo bispectrum on the choice of the mass function.
For the Press-Schechter (PS) mass function given by
\begin{eqnarray}
f_{\rm MF}(\nu) = f_{\rm PS}(\nu) = \sqrt{2 \over \pi} \nu e^{-\nu^2/2},
\label{eq:PSMF}
\end{eqnarray}  
 we have the Lagrangian bias parameter as
 \begin{eqnarray}
 b_1^\rL = {\nu^2 - 1 \over \delta_c},~b_2^\rL = {\nu^4 - 3 \nu^2 \over \delta_c^2}, ~\cdots.
 \end{eqnarray}
For the high peak formalism (HP) corresponding the large $\nu = \delta_c / \sigma_M$ limit,
we have 
\begin{eqnarray}
b_1^\rL = {\nu^2 \over \delta_c},~
b_2^\rL = {\nu^4 \over \delta_c^2},~
b_3^\rL = {\nu^6 \over \delta_c^3}, \cdots
\label{eq:HPMF}
\end{eqnarray}
The Sheth-Tormen fitting formula (ST) is given by
\begin{eqnarray}
f_{\rm MF} (\nu) = f_{\rm ST}(\nu) = A( p ) \sqrt{{2 \over \pi}}
\left[ 1 + (q\nu^2)^{-p} \right] \sqrt{q} \nu e^{-q\nu^2/2},
\label{eq:STMF}
\end{eqnarray}
and the MICE mass function (MICE) \cite{Crocce:2009mg} is given by
\begin{eqnarray}
f_{\rm MF} (\nu) = f_{\rm MICE}(\nu) = 
A_{\rm MICE} \left[ \left({\nu \over \delta_c} \right)^a + b\right] e^{-c \nu^2 / \delta_c^2},
\label{eq:MICEMF}
\end{eqnarray}
with $A_{\rm MICE}(z)= 0.58 (1+z)^{-0.13}$, $a(z)=1.37(1+z)^{-0.15}$, $b(z)=0.3 (1+z)^{-0.084}$, and $c(z)=1.036(1+z)^{-0.024}$.
In Fig. \ref{fig:fgtnlmf.eps},
we plot each contribution of the halo bispectrum, $B_{\rm grav}$ (black), $B_{\rm bis}^{\rm tree}$ (red thick), $B_{\rm bis}^{{\rm loop},1}$ (red thick dashed),
   and $B_{\rm bis}^{{\rm loop},2}$ (red thick dotted) in top panels, and $B_{\rm grav}$ (black), $B_{g_{\rm NL}}$ (blue thick), $B_{\tau_{\rm NL}}$ (green thick),
   and $B_{\rm tris}^{{\rm loop},3}$ (green thick dashed) in bottom panels, with the scale-independent bias parameters given by Eq. (\ref{eq:HPMF}) (HP),
   and the mass function given by (\ref{eq:PSMF}) (PS), (\ref{eq:STMF}) (ST) and (\ref{eq:MICEMF}) (MICE) from left to right.
   For the halo with mass $M = 5 \times 10^{13} h^{-1} M_\odot$ at $z=1.0$, we have $\nu=2.68$.
In this figure, we find that the amplitude of the halo bispectrum is also strongly dependent on the choice of the mass function.


\begin{thebibliography}{99}

\bibitem{Ade:2013ydc} 
  P.~A.~R.~Ade {\it et al.}  [Planck Collaboration],
  arXiv:1303.5084 [astro-ph.CO].
 
 
   
\bibitem{Slosar:2008hx} 
  A.~Slosar, C.~Hirata, U.~Seljak, S.~Ho and N.~Padmanabhan,
  JCAP {\bf 0808}, 031 (2008)
  [arXiv:0805.3580 [astro-ph]].
  
\bibitem{Giannantonio:2013uqa} 
  T.~Giannantonio, A.~J.~Ross, W.~J.~Percival, R.~Crittenden, D.~Bacher, M.~Kilbinger, R.~Nichol and J.~Weller,
  arXiv:1303.1349 [astro-ph.CO].

 
\bibitem{Dalal:2007cu} 
  N.~Dalal, O.~Dore, D.~Huterer and A.~Shirokov,
  Phys.\ Rev.\ D {\bf 77}, 123514 (2008)
  [arXiv:0710.4560 [astro-ph]].
  
\bibitem{Matarrese:2008nc} 
  S.~Matarrese and L.~Verde,
  Astrophys.\ J.\  {\bf 677}, L77 (2008)
  [arXiv:0801.4826 [astro-ph]].
 
 
\bibitem{Jeong:2009vd} 
  D.~Jeong and E.~Komatsu,
  Astrophys.\ J.\  {\bf 703}, 1230 (2009)
  [arXiv:0904.0497 [astro-ph.CO]].
  
  

\bibitem{Sefusatti:2007ih}
  E.~Sefusatti and E.~Komatsu,
  Phys.\ Rev.\ D {\bf 76} (2007) 083004
  [arXiv:0705.0343 [astro-ph]].
 

\bibitem{Sefusatti:2009qh} 
  E.~Sefusatti,
  Phys.\ Rev.\ D {\bf 80}, 123002 (2009)
  [arXiv:0905.0717 [astro-ph.CO]].
  
 
  
\bibitem{Baldauf:2010vn} 
  T.~Baldauf, U.~Seljak and L.~Senatore,
  JCAP {\bf 1104}, 006 (2011)
  [arXiv:1011.1513 [astro-ph.CO]].

  
  
  
\bibitem{Nishimichi:2009fs} 
  T.~Nishimichi, A.~Taruya, K.~Koyama and C.~Sabiu,
  JCAP {\bf 1007}, 002 (2010)
  [arXiv:0911.4768 [astro-ph.CO]].
  

\bibitem{Abbott:2005bi} 
  T.~Abbott {\it et al.}  [Dark Energy Survey Collaboration],
  astro-ph/0510346.

\bibitem{Schlegel:2011zz} 
  D.~Schlegel {\it et al.}  [BigBoss Experiment Collaboration],
  arXiv:1106.1706 [astro-ph.IM].


\bibitem{Abell:2009aa} 
  P.~A.~Abell {\it et al.}  [LSST Science and LSST Project Collaborations],
  arXiv:0912.0201 [astro-ph.IM].
  
\bibitem{Laureijs:2011gra} 
  R.~Laureijs {\it et al.}  [EUCLID Collaboration],
  arXiv:1110.3193 [astro-ph.CO].
  
\bibitem{Ellis:2012rn} 
  R.~Ellis {\it et al.}  [PFS Team Collaboration],
  arXiv:1206.0737 [astro-ph.CO].

 
\bibitem{Matsubara:2011ck} 
  T.~Matsubara,
  Phys.\ Rev.\ D {\bf 83}, 083518 (2011)
  [arXiv:1102.4619 [astro-ph.CO]].


\bibitem{Matsubara:2012nc} 
  T.~Matsubara,
  Phys.\ Rev.\ D {\bf 86}, 063518 (2012)
  [arXiv:1206.0562 [astro-ph.CO]].

\bibitem{Yokoyama:2012az} 
  S.~Yokoyama and T.~Matsubara,
  Phys.\ Rev.\ D {\bf 87}, 023525 (2013)
  [arXiv:1210.2495 [astro-ph.CO]].
 
 
\bibitem{Sato:2013qfa} 
  M.~Sato and T.~Matsubara,
  arXiv:1304.4228 [astro-ph.CO].
 
 
  
\bibitem{Hinshaw:2012aka} 
  G.~Hinshaw {\it et al.}  [WMAP Collaboration],
  Astrophys.\ J.\ Suppl.\  {\bf 208}, 19 (2013)
  [arXiv:1212.5226 [astro-ph.CO]].

  
\bibitem{Bernardeau:2008fa} 
  F.~Bernardeau, M.~Crocce and R.~Scoccimarro,
  Phys.\ Rev.\ D {\bf 78}, 103521 (2008)
  [arXiv:0806.2334 [astro-ph]].
  
 
\bibitem{Matsubara:2013ofa} 
  T.~Matsubara,
  arXiv:1304.4226 [astro-ph.CO].
  
  
\bibitem{Suyama:2007bg} 
  T.~Suyama and M.~Yamaguchi,
  Phys.\ Rev.\ D {\bf 77}, 023505 (2008)
  [arXiv:0709.2545 [astro-ph]].
  
\bibitem{Suyama:2010uj} 
  T.~Suyama, T.~Takahashi, M.~Yamaguchi and S.~Yokoyama,
  JCAP {\bf 1012}, 030 (2010)
  [arXiv:1009.1979 [astro-ph.CO]].
 
\bibitem{Sugiyama:2011jt} 
  N.~S.~Sugiyama, E.~Komatsu and T.~Futamase,
  Phys.\ Rev.\ Lett.\  {\bf 106}, 251301 (2011)
  [arXiv:1101.3636 [gr-qc]].
 
\bibitem{Bramante:2011zr} 
  J.~Bramante and J.~Kumar,
  JCAP {\bf 1109}, 036 (2011)
  [arXiv:1107.5362 [astro-ph.CO]].
 
\bibitem{Sugiyama:2012tr} 
  N.~S.~Sugiyama,
  JCAP {\bf 1205}, 032 (2012)
  [arXiv:1201.4048 [gr-qc]].
  
  



\bibitem{Tseliakhovich:2010kf} 
  D.~Tseliakhovich, C.~Hirata and A.~Slosar,
  Phys.\ Rev.\ D {\bf 82}, 043531 (2010)
  [arXiv:1004.3302 [astro-ph.CO]].
  
\bibitem{Smith:2010gx} 
  K.~M.~Smith and M.~LoVerde,
  JCAP {\bf 1111}, 009 (2011)
  [arXiv:1010.0055 [astro-ph.CO]].
  
  

\bibitem{Gong:2011gx} 
  J.~-O.~Gong and S.~Yokoyama,
  arXiv:1106.4404 [astro-ph.CO].
 
\bibitem{Nishimichi:2012da} 
  T.~Nishimichi,
  JCAP {\bf 1208}, 037 (2012)
  [arXiv:1204.3490 [astro-ph.CO]].
 
\bibitem{Biagetti:2012xy} 
  M.~Biagetti, V.~Desjacques and A.~Riotto,
  arXiv:1208.1616 [astro-ph.CO].
 
 
\bibitem{Baumann:2012bc} 
  D.~Baumann, S.~Ferraro, D.~Green and K.~M.~Smith,
  JCAP {\bf 1305}, 001 (2013)
  [arXiv:1209.2173 [astro-ph.CO]].
  
  
    
\bibitem{Sheth:1999mn} 
  R.~K.~Sheth and G.~Tormen,
  Mon.\ Not.\ Roy.\ Astron.\ Soc.\  {\bf 308}, 119 (1999)
  [astro-ph/9901122].

  
  
  
\bibitem{Mizuno:2010by} 
  S.~Mizuno and K.~Koyama,
  JCAP {\bf 1010}, 002 (2010)
  [arXiv:1007.1462 [hep-th]].


\bibitem{Izumi:2011di} 
  K.~Izumi, S.~Mizuno and K.~Koyama,
  Phys.\ Rev.\ D {\bf 85}, 023521 (2012)
  [arXiv:1109.3746 [astro-ph.CO]].
  
  
\bibitem{Sefusatti:2011gt} 
  E.~Sefusatti, M.~Crocce and V.~Desjacques,
  arXiv:1111.6966 [astro-ph.CO].

  
\bibitem{Crocce:2009mg} 
  M.~Crocce, P.~Fosalba, F.~J.~Castander and E.~Gaztanaga,
  Mon.\ Not.\ Roy.\ Astron.\ Soc.\  {\bf 403}, 1353 (2010)
  [arXiv:0907.0019 [astro-ph.CO]].

 
 
\bibitem{Matarrese:1986et} 
  S.~Matarrese, F.~Lucchin and S.~A.~Bonometto,
  Astrophys.\ J.\  {\bf 310}, L21 (1986).
 
  \end{thebibliography}
\end{document}